\documentclass[aps,prd,twocolumn,showpacs]{revtex4}

\usepackage[dvips]{graphicx}
\usepackage{xspace}
\usepackage{latexsym}
\usepackage{amssymb}
\usepackage{times}
\usepackage{mathptm}

\def\etal {{\em et al.,\ }}
\def\th13 {\theta_{13}}

\def\bc {\begin{center}}
\def\ec {\end{center}}
\def\be {\begin{equation}}
\def\bea {\begin{eqnarray}}
\def\ee {\end{equation}}
\def\eea {\end{eqnarray}}
\def\cosz {\cos\theta^{\rm zenith}}

\def\lapp{\mathrel{\rlap{\raise.5ex\hbox{$<$}}
                    {\lower.5ex\hbox{$\sim$}}}}
\def\gapp{\mathrel{\rlap{\raise.5ex\hbox{$>$}}
                    {\lower.5ex\hbox{$\sim$}}}}

\def\dam{{(\Delta m_{31}^2)^m}}
\def\da{{(\Delta m_{31}^2)}}
\begin{document}

\title{The mass hierarchy with atmospheric neutrinos at INO}

\date{\today}

\newcommand{\sinp}{\affiliation{Saha Institute of Nuclear
Physics,\\ 1/AF, Bidhannagar, Kolkata 700 064, India}}

\author {Abhijit Samanta\footnote{presently at 
Harish-Chandra Research Institute,
Chhatnag Road,
Jhusi,
Allahabad 211 019,
India.}}
\affiliation
{Saha Institute of Nuclear
Physics,\\ 1/AF, Bidhannagar, Kolkata 700 064, India}

\begin{abstract}
We study the neutrino mass hierarchy
at the magnetized Iron CALorimeter (ICAL) detector at India-based Neutrino 
Observatory with atmospheric neutrino events generated by the
Monte Carlo event generator Nuance.
We judicially choose  the observables so that the possible systematic uncertainties 
can be reduced. 
The resolution as a function of both energy and zenith angle { simultaneously}
is obtained for neutrinos and anti-neutrinos separately
from thousand years un-oscillated atmospheric neutrino events at ICAL
{ to migrate number of events from neutrino energy and zenith angle bins to muon
energy and zenith angle bins}. 
The resonance ranges in terms of { directly measurable quantities like}
muon energy and zenith angle are found
using this resolution function at different input values of $\theta_{13}$.  
Then, the marginalized $\chi^2$s are studied for different input values of 
$\theta_{13}$ 
with its resonance ranges { taking input data in muon energy and zenith 
angle bins}.
Finally, we find that the mass hierarchy can be explored up to a lower value 
of $\theta_{13}\approx 5^\circ$ { with confidence level $>$ 95\%} in this set up.   

\end{abstract}



\maketitle

\section{introduction}

The recent experiments reveal that neutrinos have mass and they 
oscillate \cite{pdg} from one flavor to another flavor as they travel.
In the standard framework of oscillation scenario with three active 
neutrinos,
the  mass eigenstates $|\nu_i\rangle$ and the flavor eigen states 
$|\nu_\alpha\rangle$ are related by a mixing matrix $U$ \cite{pmns}: 
\be |\nu_\alpha\rangle = U_{\alpha i} |\nu_i\rangle. \ee
The flavor mixing matrix in vacuum $U$ (PDG representation \cite{pdg}) is
\bea
\left (
\begin{array}{c c c}
c_{12} c_{13} & s_{12} c_{13} & s_{13} e^{-i\delta}\\
-s_{12}c_{23}-c_{12}s_{23}s_{13}e^{i\delta} &
c_{12}c_{23}-s_{12}s_{23}s_{13}e^{i\delta} & s_{23}c_{13}\\
s_{12}s_{23}-c_{12}c_{23}s_{13}e^{i\delta} &
-c_{12}s_{23}-s_{12}c_{23}s_{13}e^{i\delta} & c_{23}c_{13}
\end{array}
\right ).
\label{e:u}
\eea

In this oscillation picture, there are six parameters: two mass squared 
differences $\Delta m_{ij}^2$ ($\Delta m_{12}^2, \Delta m_{32}^2$),
three mixing angles ($\theta_{12}, \theta_{23}, \theta_{13}$), and a
CP-violating phase $\delta$. 
At present, information is available on two neutrino mass-squared differences
and two mixing angles: from atmospheric neutrinos one gets the best-fit values with
$3\sigma$ error
$|\Delta m^{2}_{32}|\simeq 2.50^{+.52}_{-0.60}\times 10^{-3}$
eV$^2$, $\theta_{23}\simeq$ ${45.0^\circ}^{+10.55^\circ}_{-9.44^\circ}$;
while solar neutrinos
tell us $\Delta m^{2}_{21} \simeq 7.9\times 10^{-5}$ eV$^2$,
$\theta_{12}\simeq {33.21^\circ}^{+6.02^\circ}_{-3.88^\circ}$ \cite{Schwetz:2006dh}.
{Here $\Delta m^{2}_{ij}$= $m^{2}_{i} - m^{2}_{j}$.}
The present experiments are sensitive to $\left|\Delta m_{32}^2\right|$ 
only and the both sign of $\Delta m_{32}^2$ fit the data equally 
well.

The neutrino mass hierarchy, whether normal ($m_3^2 > m_2^2$), or
inverted ($m_2^2 > m_3^2$), is of great theoretical interest. 
For example, the grand unified theory favors normal hierarchy.
It can be qualitatively understood from the fact that it relates
leptons to quarks and quark mass hierarchies are normal.
The inverted mass hierarchy which is unquarklike, would 
probably indicate a global leptonic symmetry\cite{Mohapatra:2004vr}. 

For nonzero value of $\theta_{13}$, the hierarchy can be determined 
from the matter effect on neutrino oscillation. 
The contributions of this 
effect  
to the effective  Lagrangian during the propagation through  matter 
are opposite for neutrino and anti-neutrino, which depends mainly on 
the value of $\theta_{13}$ and the sign of $\Delta m_{32}^2$.
So the  total number of observed events is expected to differ 
for normal hierarchy (NH) and inverted hierarchy (IH). The hierarchy can be more prominently
distinguished if the experiment is able to count neutrinos and anti-neutrinos
separately. This needs charge identification of the produced leptons.

In case of $\theta_{13}=0$, the mass hierarchy can also be observable
in principle even in case of vacuum oscillation due to nonzero value of 
$\Delta m_{21}^2$\cite{deGouvea:2005hk}.

Presently over the world, there are many ongoing and planned experiments:
MINOS\cite{Zois:2004ns}, T2K\cite{Yamada:2006hi}, ICARUS\cite{Kisiel:2005ti},
NOVA\cite{Ray:2006ke}, D-CHOOZ\cite{Horton-Smith:2006yh},
UNO\cite{Jung:1999jq}, SKIII\cite{Back:2004qi}, OPERA\cite{DiCapua:2005bd},
Hyper-K\cite{Nakamura:2003hk}, and many  others. It is notable that all these experiments are 
planned in the north hemisphere of the Earth. 
Among them MINOS is the only magnetized one and has a good charge 
identification capability. A large magnetized ICAL
detector is under strong consideration for the proposed India-based
Neutrino Observatory (INO)\cite{ino} near the equator.
Since it has high charge identification capability $(\gapp 95\%)$
\cite{Arumugam:2005nt}, it is expected to have reasonably better capability 
to determine the neutrino 
mass hierarchy. 

It is also notable here that the CERN-INO baseline happens to be
close to the so-called `magic' baseline \cite{magic,Agarwalla:2005we}
for which the oscillation probabilities are relatively insensitive to the yet
unconstrained CP phase.  This permits such an experiment to make precise
measurements of the masses and their mixing avoiding the degeneracy issues
\cite{degeneracy} which are associated with other baselines.

The mass hierarchy with atmospheric neutrinos at the magnetized ICAL  
has been studied in \cite{Petcov:2005rv, 
Gandhi:2007td,
Indumathi:2004kd}. 
In \cite{Petcov:2005rv, Gandhi:2007td},  the number of events are calculated 
in energy and zenith angle bins. Then the chi-square is calculated
including the effect of possible uncertainties over a large range of 
zenith angle and energy. 
It was found in \cite{Petcov:2005rv} that a precision of $\simeq 5\%$ in 
neutrino energy  
and 5$^\circ$ in neutrino direction reconstruction  are required to 
distinguish the hierarchy at 2$\sigma$ level with 200 events for 
$\sin^22\theta_{13}=0.1$ and roughly an order of  magnitude larger event numbers are required in case of resolutions of 15\% for the neutrino energy and $15^\circ$ for 
the neutrino direction. However, in \cite{Gandhi:2007td}, the authors
are able to relax the resolution width to $10-15\%$ for energy and $10^\circ$ 
for direction reconstruction in distinguishing 
the type of hierarchy at 95\% CL or better for $\sin^22\theta_{13} \ge 0.05$
with 1 Mton-yr exposure of ICAL.

In \cite{Indumathi:2004kd},  an asymmetry parameter is considered as 

\be{\cal{A}}_N(L/E) = \frac{u}{d}(L/E)-\frac{\overline{u}}{\overline{d}}(L/E)
\ee
where, $u,~(\bar u),~d,~(\bar d)$ are the number of up going and down going 
events for neutrino (anti-neutrino), respectively for a baseline $L$
and an energy $E$.
For down going events,  the `mirror' $L$ is considered, which is exactly
equal to that value when neutrino comes from the opposite direction. 
Here the asymmetry is calculated  integrating over $E
 \geq 4$ GeV. Here a statistically significant region with maximum 
number of events corresponds to the range $500 \lapp L/E \lapp 3000$.
However, the results depends crucially on the $L/E$ resolutions.

In \cite{Gandhi:2005wa}, the neutrino and the anti-neutrino events for direct as well as 
inverted hierarchy are considered for the $E$ range $5 - 10$ GeV and
the $L$ range $6000 - 9500$ km.  

In our work, we find  the resonance ranges in terms of directly 
measurable parameters $\theta_\mu^{\rm zenith}$ and $E_\mu$ for a given set of oscillation 
parameters, where the matter effect contributes significantly in distinguishing 
the hierarchy. 
We have also studied the possible observables 
and find the suitable ones  for discrimination of the hierarchy. Finally, 
we make a marginalized $\chi^2$ study over a range of oscillation parameters 
and find how low $\theta_{13}$ can be explored in discriminating the hierarchy
at INO.   

\section{The INO detector}
The detector is  a rectangular parallelepiped magnetized Iron CALorimeter
detector\cite{ino}. Here the simulation has been carried out for 50 kTon 
mass with size 48 m $\times$ 16 m $\times$ 12 m. It consists of a 
stack of 140 horizontal layers of 6 cm thick iron slabs interleaved with 
2.5 cm gap for the active detector elements. For the sake of illustration,
we define a rectangular coordinate frame with origin at the center of the 
detector, $x(y)$-axis along the longer (shorter) lateral direction, and 
$z$-axis along the vertical direction. A magnetic field of strength $\approx$ 1 Tesla 
will be applied along  $+y$-direction. The resistive plate chamber (RPC) 
appears to be the best option for the active part of the detector. 
It is a gaseous detector consisting of  two parallel electrodes made
of 2 mm thick 2 m $\times$ 2 m glass/Bakelite plates with  graphite paint on
the out sides and separated by a gap of 2 mm.
When a charge particle passes through this active part, it gives a
transient and a very localized electric discharge in the gases.
The readout of the RPCs are the 2 cm width Cu strips placed on
the external sides of the  plates. This type of detector has a very 
good time ($\sim$ 1 ns) as well as spatial resolution.

\section{Atmospheric neutrino flux and events}\label{s:flux}
The atmospheric neutrinos are produced from the interactions of the
cosmic rays with the Earth's atmosphere. The knowledge of primary spectrum 
of the cosmic rays has been  improved from the observations by 
BESS\cite{Maeno:2000qx} and AMS\cite{Alcaraz:2000ss}. However  large regions 
of parameter space have been unexplored and they are interpolated 
or extrapolated from the measured flux.  The difficulties and uncertainties
in calculation of the neutrino flux depends on the neutrino energy.
The low energy fluxes have been known quite well. The 
cosmic ray fluxes ($<$ 10 GeV)  are modulated by solar activity
and geomagnetic field through a rigidity (momentum/charge) cutoff.
At higher neutrino energy ($>$ 100 GeV), solar activity and rigidity 
cutoff are irrelevant\cite{Honda:2004yz}.  
There is 10\% agreement among the calculations for neutrino energy below 10 GeV
because different hadronic interaction models are used in the
calculations and because 
the uncertainty in cosmic ray flux measurement
is 5\% for cosmic ray energy below 100 GeV \cite{Honda:2004yz}. 
In our simulation, we have used a typical Honda flux calculated in 
3-dimensional scheme\cite{Honda:2004yz}. 

The  interactions of neutrinos with the detector material are simulated 
using the Monte Carlo model Nuance (version-3)\cite{Casper:2002sd}. 
Here the charged and 
neutral current interactions are considered for (quasi-)elastic reactions, 
resonance processes, 
coherent and diffractive,  
and deep inelastic scattering  processes.

\begin{figure*}[htb]
\includegraphics[width=6.0cm,angle=0]{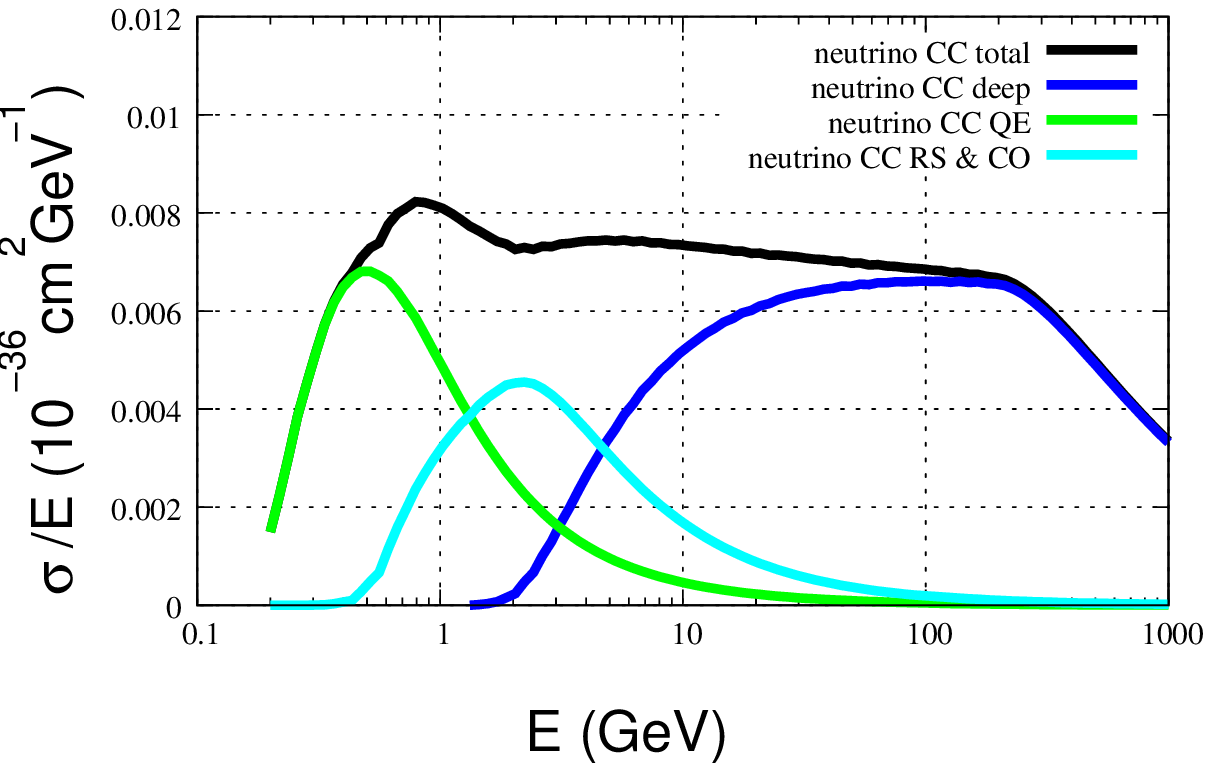}
\includegraphics[width=6.0cm,angle=0]{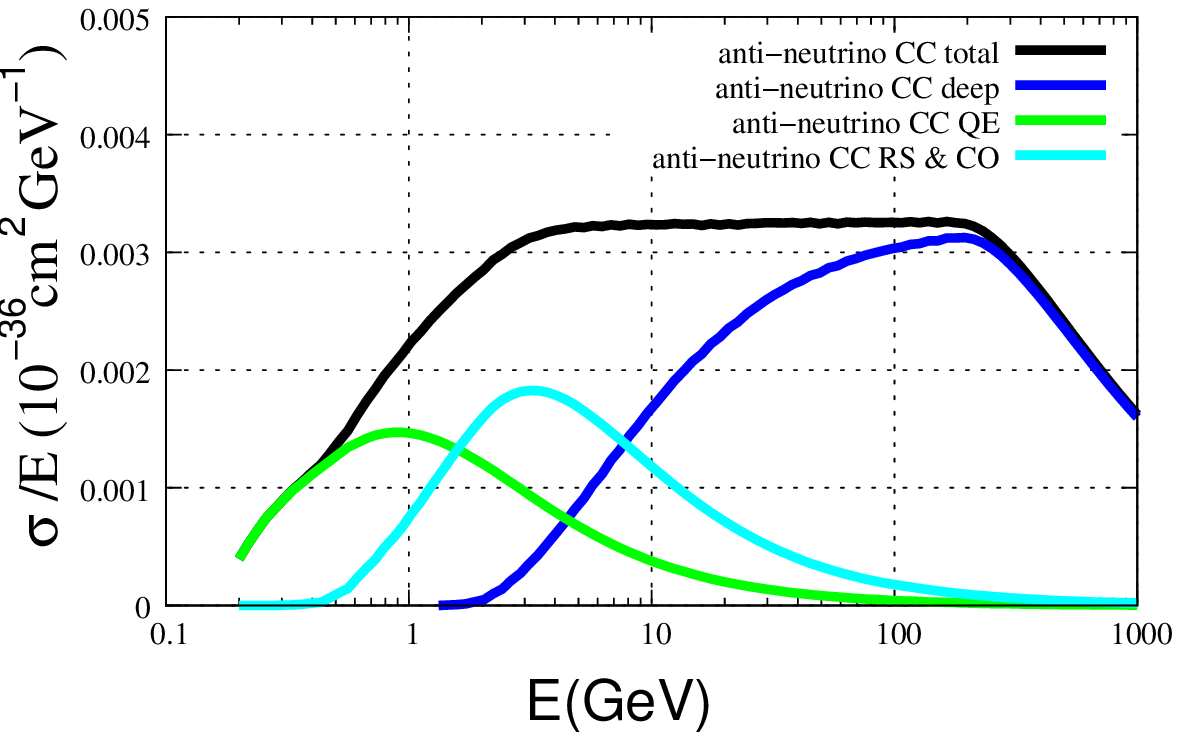}
\includegraphics[width=6.0cm,angle=270]{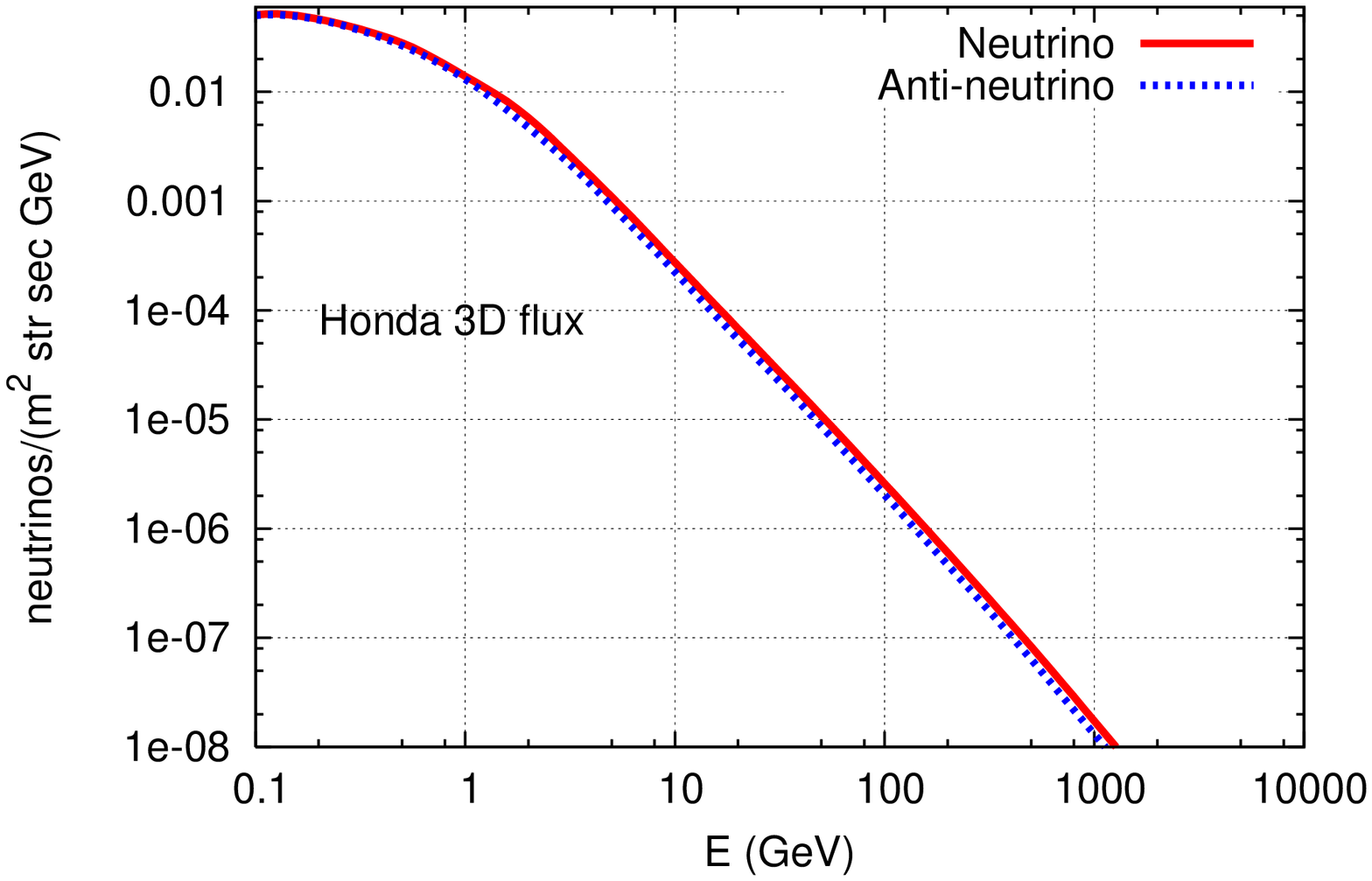}
\caption{\sf The variation of total CC iron cross section 
(scaling by a factor of 
$\frac{1}{56}$)  
with neutrino energy  for 
both neutrino (left) and anti-neutrino (right)  and the variation of atmospheric 
flux (Honda 3D) with energy (lower) for a fixed zenith angle and azimuthal angle.}
\label{f:cross}
\end{figure*}

In fig. \ref{f:cross}, we show the variation of  total Charged Current (CC) cross section 
with neutrino energy for both neutrino  and anti-neutrino. In the third plot
of this figure, we also show how rapidly the neutrino flux changes with energy. 

\section{Neutrino oscillation through the Earth matter}\label{s:oscillation}
The effective Hamiltonian which describes the time evolution of neutrinos in
matter can be expressed in flavor basis as 

\bea
H=\frac{1}{2 E}\left\{U
\left(\begin{array}{c c c} 0&&\\&\Delta m_{21}^2&\\&&\Delta m_{31}^2\end{array}\right)
U^+  +
\left(\begin{array}{c c c} A&&\\&0&\\&&0\end{array}\right) \right\}
\eea
where, $A=G\sqrt{2}N_e 2E$ is the  
effective potential of $\nu_e$ with electrons, 
$U$ is the flavor mixing matrix in vacuum  (eq. \ref{e:u}),
$G$ is the Fermi constant, $N_e$ is the electron density 
of the medium, and $E$ is the neutrino energy.


For a time t, the evolution of neutrino states is given by
$\nu(t)=S(t) \nu(0)$
with
$S(t)=e^{-i H t}$
for constant matter density. The corresponding effective Hamiltonian
for anti-neutrinos is obtained by $U \rightarrow U^*$ and
$A \rightarrow -A$.
To understand the analytical solution one may adopt the 
so called ``one mass scale dominance" (OMSD) frame work: 
$|\Delta m_{21}^2| << | m_{3}^2 -m_{1,2}^2|$
\cite{Peres:2003wd,Akhmedov:2004ny}.

With this OMSD approximation, the survival probability of $\nu_\mu$
can expressed as
\bea
P^{m}_{\mu \mu} &=&
{1 - \cos^2 \theta^m_{13} \; {\sin^2 2 \theta_{23}}} 
\nonumber\\
&&
\times\sin^2\left[1.27 \;\left(\frac{\da + A + \dam}{2}\right) \;\frac{L}{E} \right]
\nonumber \\
&& ~-~
\sin^2 \theta^m_{13}\; \sin^2 2 \theta_{23}
\nonumber \\
&& 
\times\sin^2\left[1.27 \;\left(\frac{\da + A - \dam}{2}\right) \;\frac{L}{E}
\right]
\nonumber \\
&& ~-~
{{\sin^4 \theta_{23}}} \;
{ { \sin^2 2\theta^m_{13} \;
\sin^2 \left[1.27\; \dam  \;\frac{L}{E} \right]
}}
\label{e:pmumu}
\eea

 The  mass squared difference ${{\dam}}$ and mixing angle
${ {\sin^22\theta_{13}^m}}$ in matter are related to their vacuum values by
\bea
\dam =
\nonumber
\sqrt{(\da \cos 2 \theta_{13} - A)^2 +
(\da \sin 2 \theta_{13})^2} 
\\
sin2\theta^m_{13}=
\frac{{\da \sin 2 \theta_{13}}}
{\sqrt{(\da \cos 2 \theta_{13} - A)^2 +(\da \sin 2 \theta_{13})^2.} }
\label{e:dm31}
\eea
 
From eq. \ref{e:pmumu} and \ref{e:dm31} it is seen that a
resonance in ${{P^{m}_{\mu \mu} }}$ will occur for 
neutrinos (anti-neutrinos) with NH (IH) when
\be \sin^22\theta_{13}^m \rightarrow 1
 ~~~~{\rm or,}~~ A= \Delta m^2_{31} \cos2\theta_{13}. \ee

Then resonance energy can be expressed as 
\bea E = \left [\frac{1}{2\times 0.76 \times 10^{-4} Y_e} \right]
\left [\frac{| \Delta m_{31}^2|}{\rm eV^2} \cos2\theta_{13} \right ]
\left[\frac {\rm gm/cc}{\rho} \right ]
\label{e:resonance}\eea

\begin{figure*}[htb]
\includegraphics[width=8.0cm,angle=270]{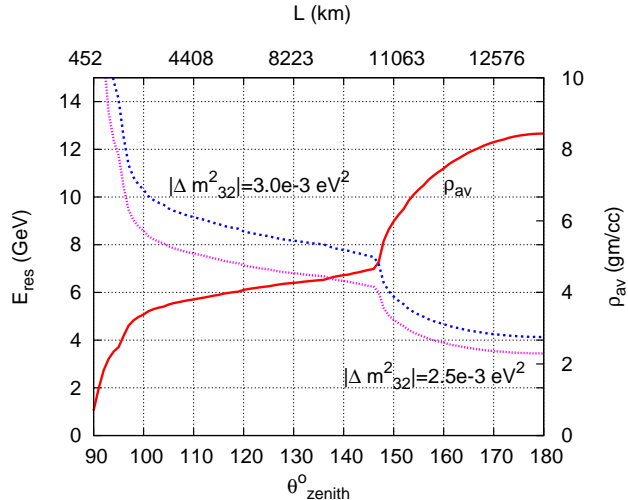}
\caption{\sf The variation of average density (y2-axis) and the corresponding
resonance energy (y1-axis) with zenith angle (x1-axis) and baseline (x2-axis) 
at $13$ resonance assuming one mass scale dominance
approximation. We set $\theta_{13}= 10^\circ$. }
\label{f:resonance}
\end{figure*}

Fig. \ref{f:resonance} shows  how the resonance 
energy and the average density of the Earth changes with zenith angle.

\section{Role of hierarchy in oscillation probability at different 
{ $E_\nu$ and $\cosz_\nu$}}\label{s:role}
\begin{figure*}[htb]
\includegraphics[width=5.0cm,angle=270]{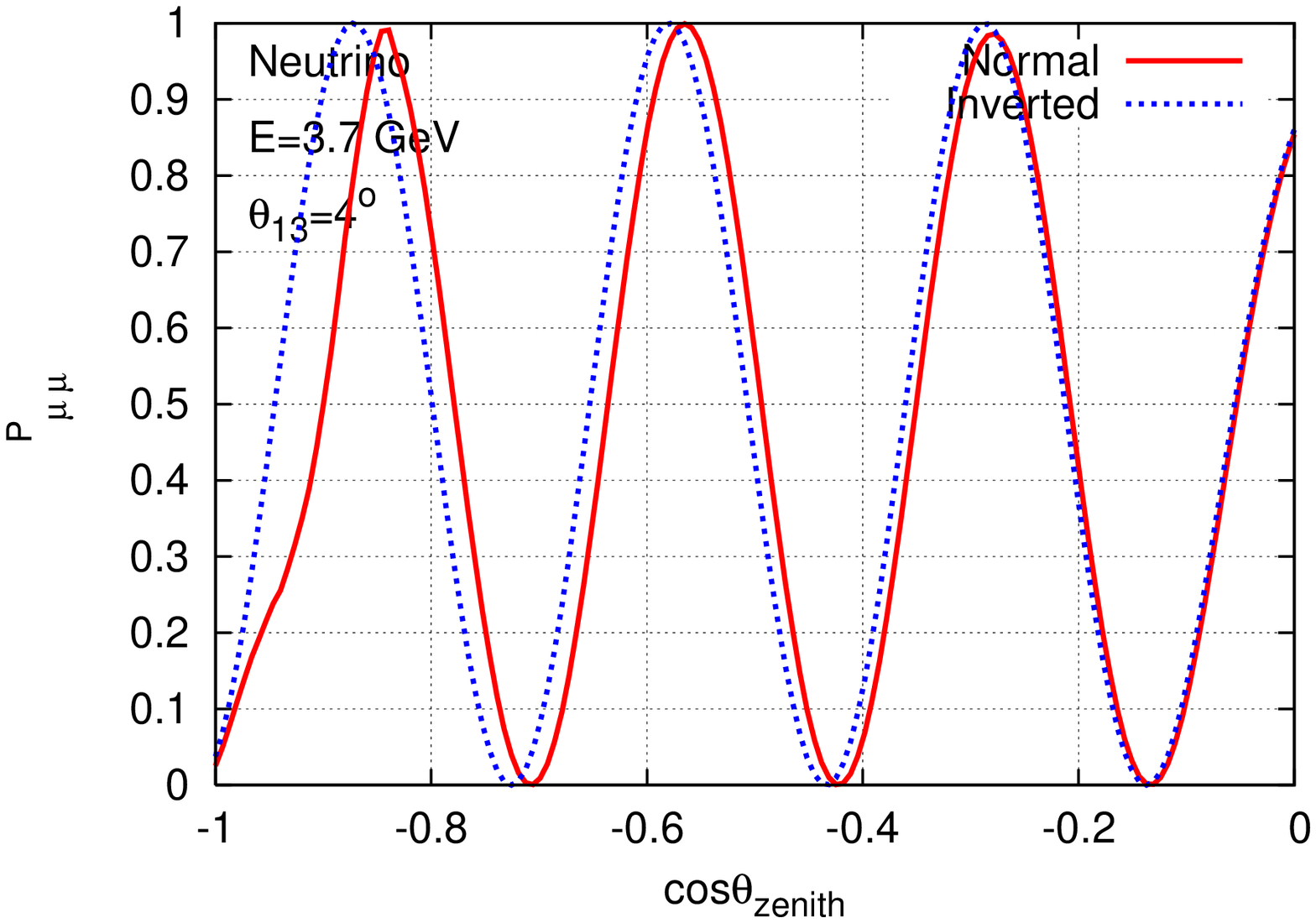}
\includegraphics[width=5.0cm,angle=270]{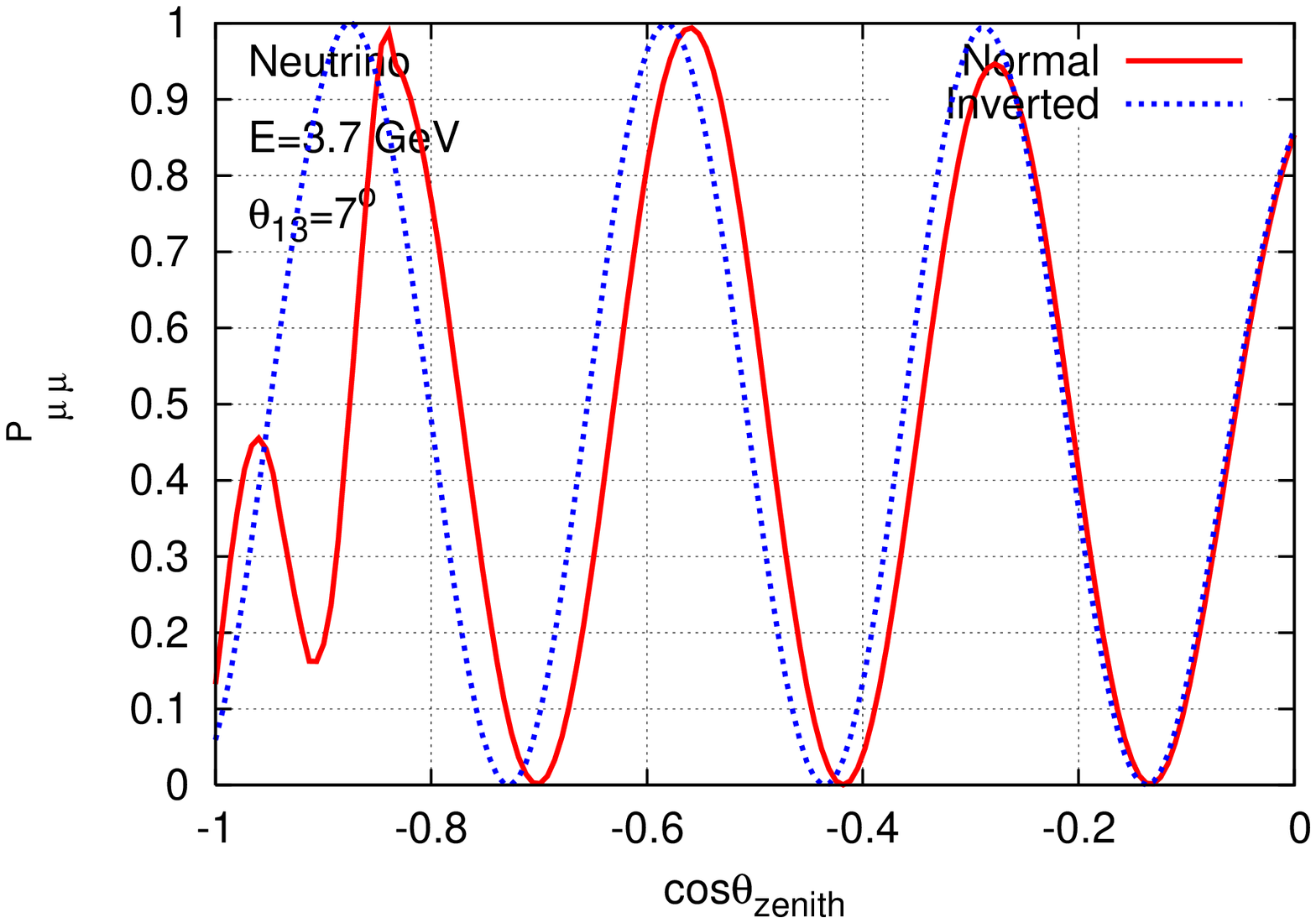}
\includegraphics[width=5.0cm,angle=270]{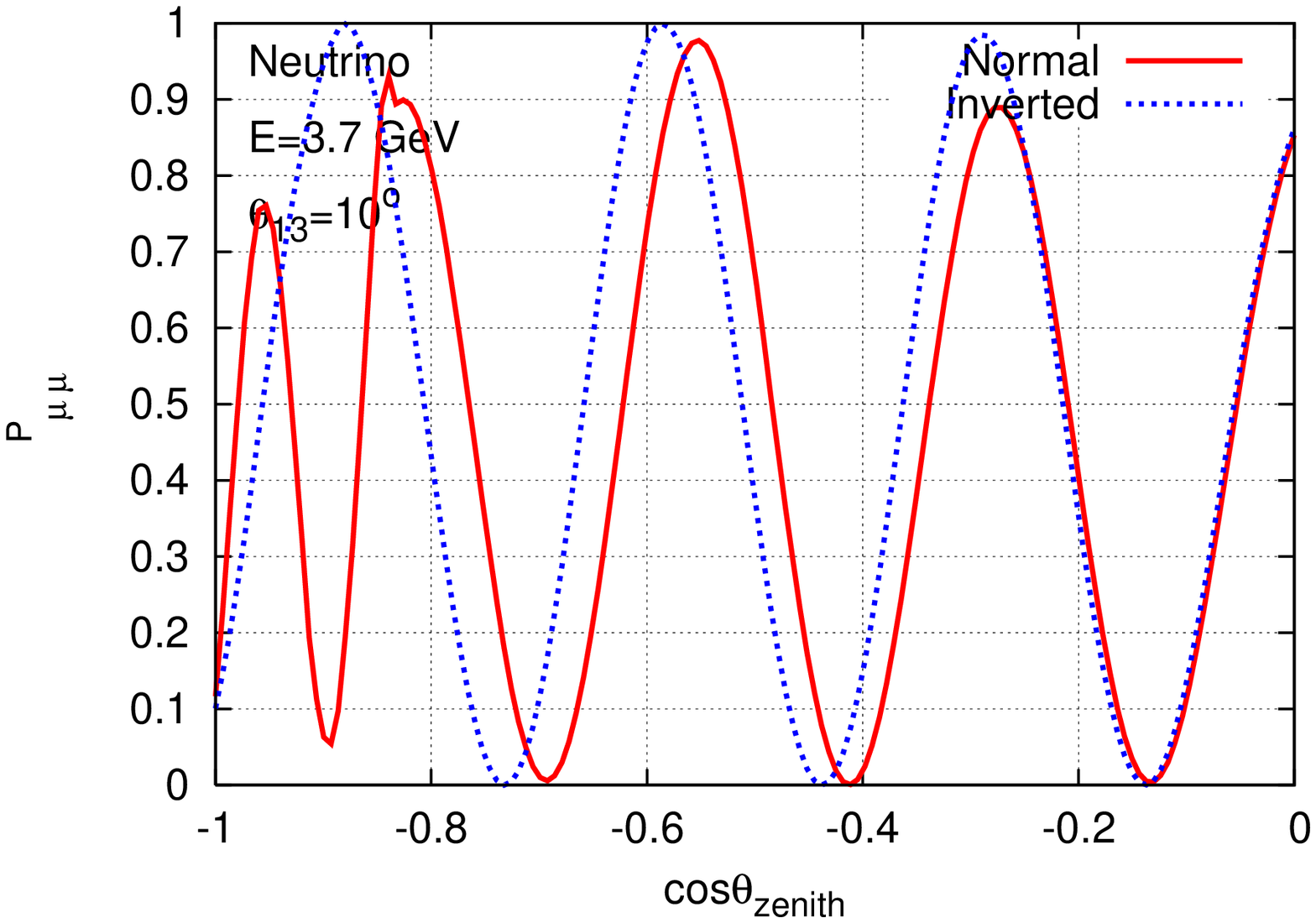}
\includegraphics[width=5.0cm,angle=270]{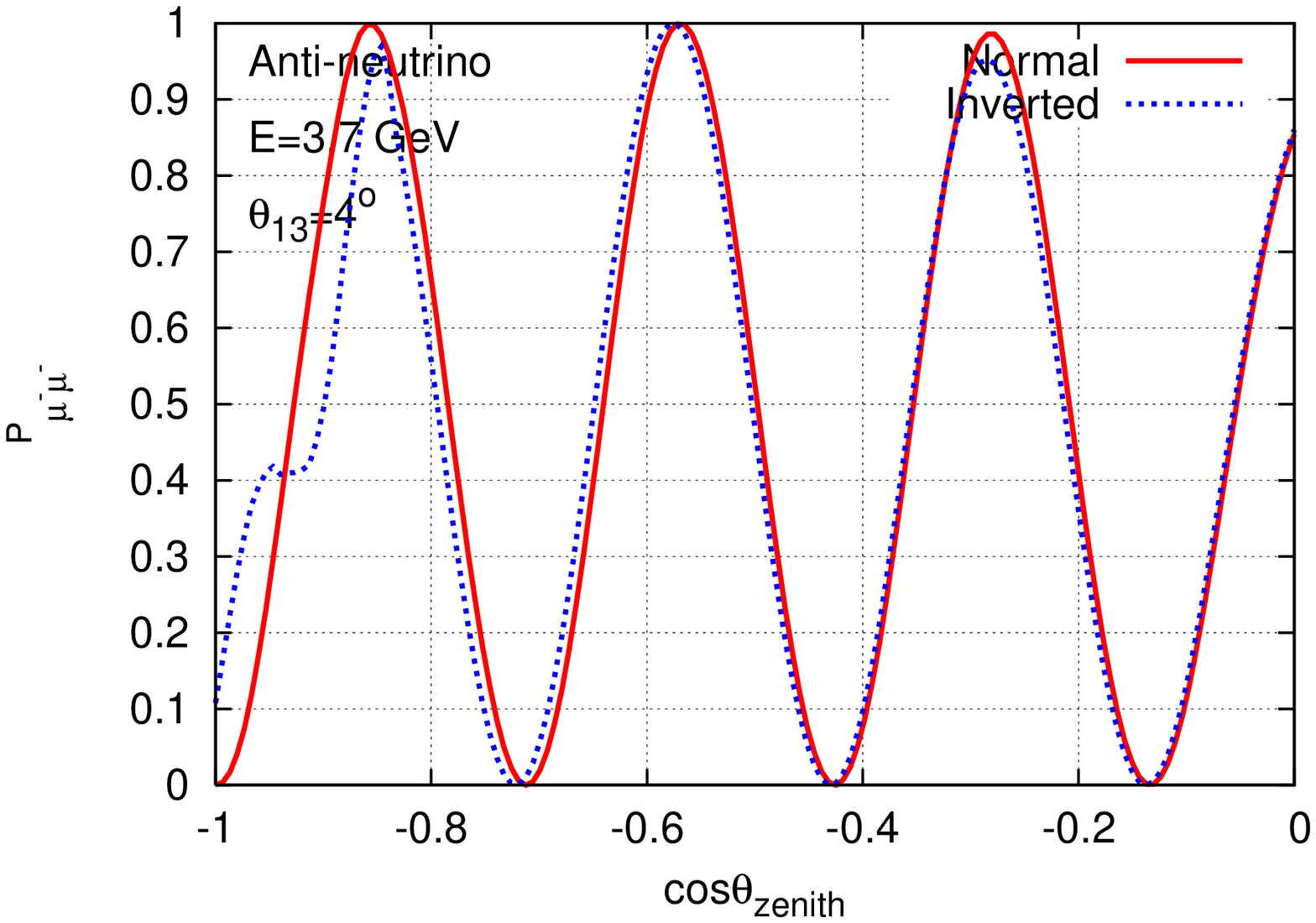}
\includegraphics[width=5.0cm,angle=270]{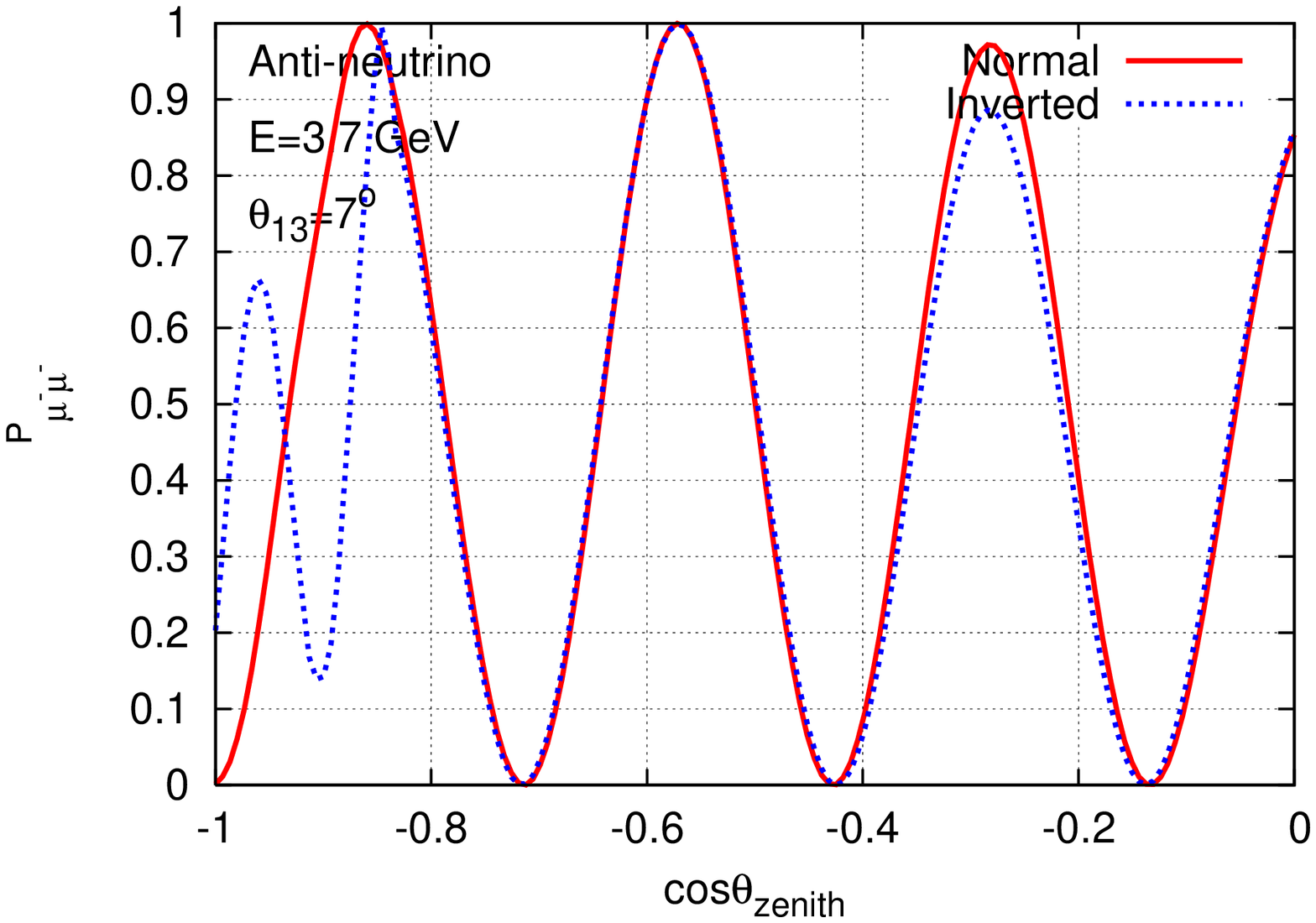}
\includegraphics[width=5.0cm,angle=270]{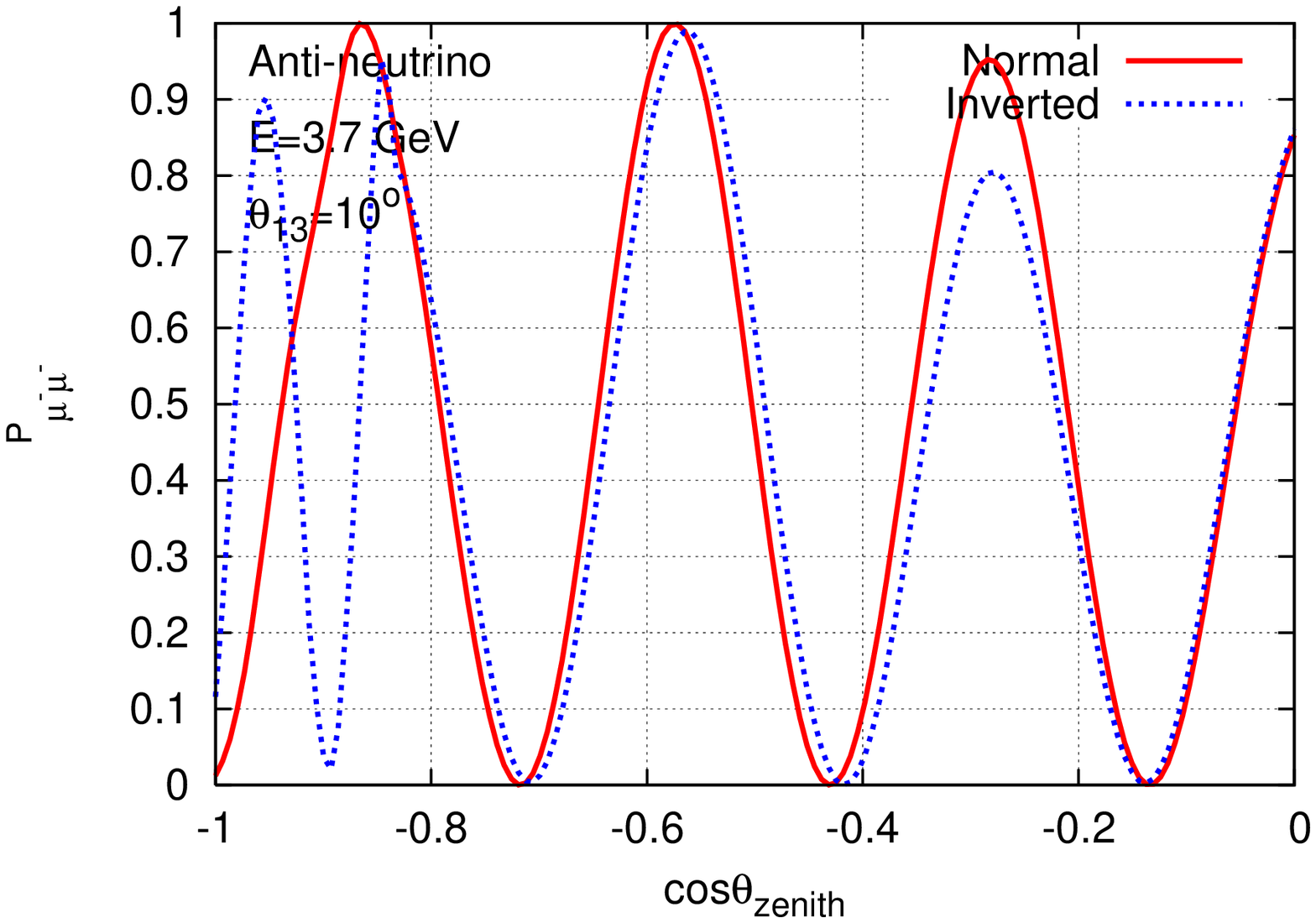}
\caption{\sf 
The variation of the $\nu_\mu$ (first row) and $\bar\nu_\mu$ (second row) survival probability
with $\cos\theta_{\rm zenith}$ with
a fixed energy $E_\nu$ = 3.7 GeV for $\theta_{13}=4^\circ$ (first column),
$7^\circ$ (second column), and $10^\circ$ (third column). 
The other oscillation parameters
are set at their best-fit values.
}
\label{f:p_nue3.7}
\end{figure*}


The significant difference in survival probabilities 
between NH and IH arises due to matter resonance.  For a neutrino 
energy $(E_{\nu})$, the corresponding baseline  
$(L_\nu)$ or zenith angle ($\theta^{\rm zenith})$ where the resonance 
occurs, can be understood from fig. \ref{f:resonance} for both cases of 
$\nu_\mu$  and $\bar\nu_\mu$. There are two zones of energy: 
\begin{enumerate}
\item low energy zone ($E\approx 4$ GeV) for the events through the {\it core} of the Earth, and
\item high energy zone ($E\approx 6 - 8$ GeV) for the events through the 
{\it mantle} of the Earth. 
\end{enumerate}
Here we have studied in detail the difference in survival probabilities 
at different  $E_{\nu}$ and $\cosz_\nu$ using full three flavor oscillation 
formula with the Earth density profile of PREM model\cite{Dziewonski:1981xy}. 
The dependence on the values of the oscillation parameters are also studied 
in detail. The observations are the following.
 
The survival probabilities for both NH and IH  without matter effect produce a sinusoidal 
behavior in $L$ for a fixed $E$.
In case of non-zero $\theta_{13}$ with NH, it is seen that when 
$\nu_\mu$ with energy $\approx 2-6$ GeV passes through the core of the 
Earth (density $\approx 12$gm/cc), a large depletion in survival 
probability $P(\nu_\mu\leftrightarrow \nu_\mu)$ with respect to IH arises 
due to the increase  of $P(\nu_\mu \leftrightarrow \nu_e)$
(see first row of fig. \ref{f:p_nue3.7}). 
In case of $\bar\nu_\mu$, it happens for IH (see second row of fig. \ref{f:p_nue3.7}).
The important point is that  the atmospheric neutrino flux is sufficiently 
high at this energy range. 

\begin{figure*}[htb]
\includegraphics[width=5.0cm,angle=270]{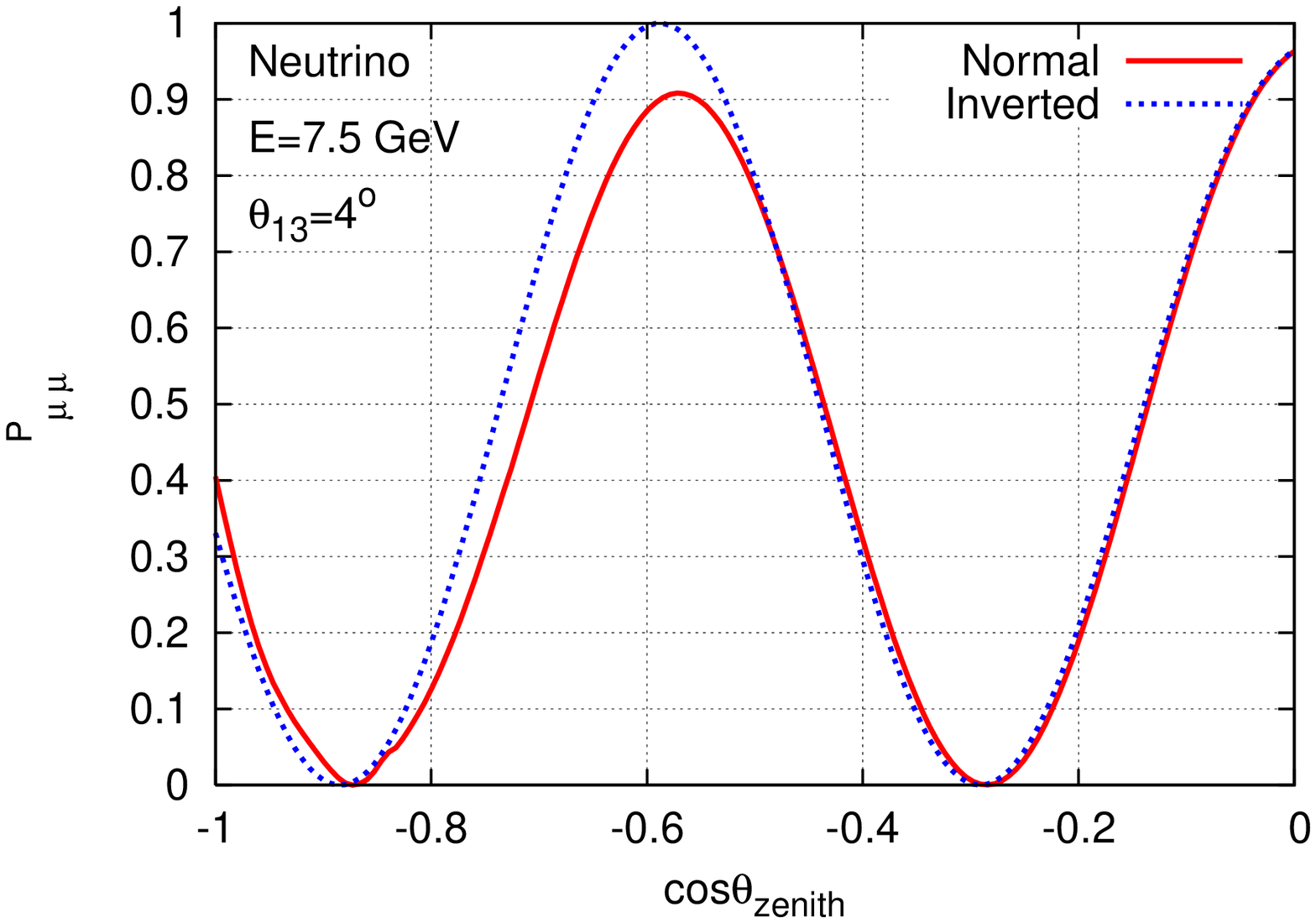}
\includegraphics[width=5.0cm,angle=270]{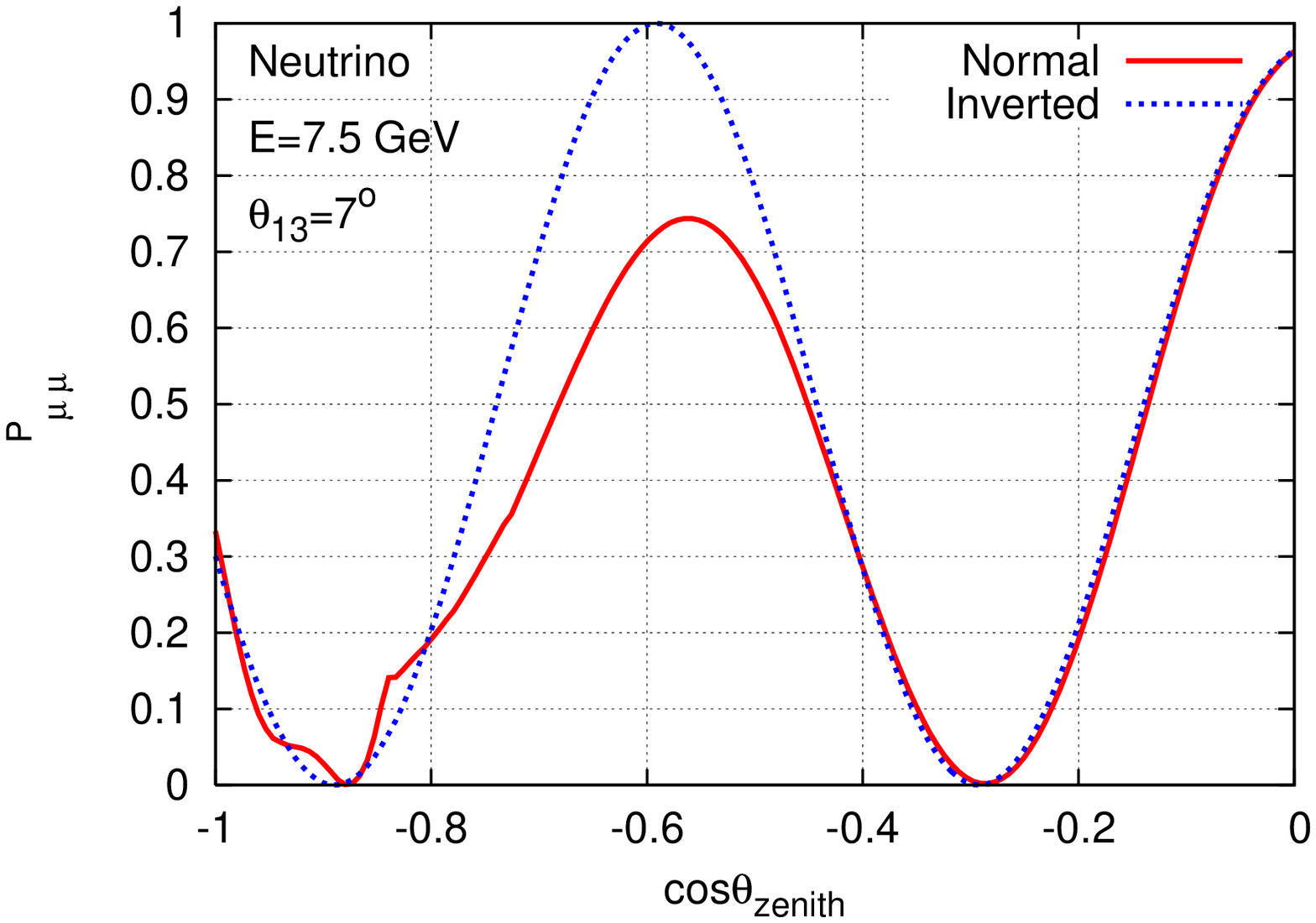}
\includegraphics[width=5.0cm,angle=270]{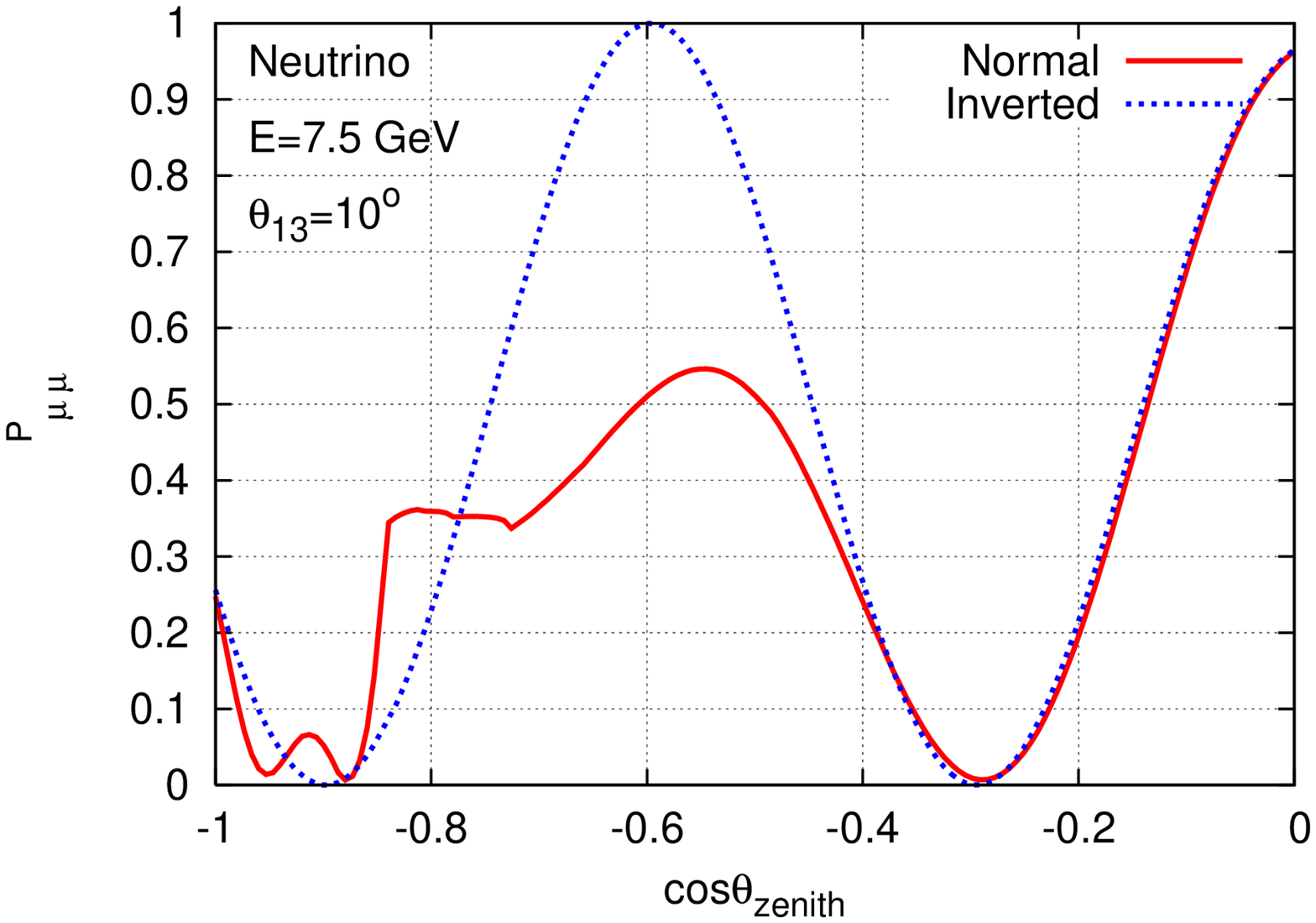}
\includegraphics[width=5.0cm,angle=270]{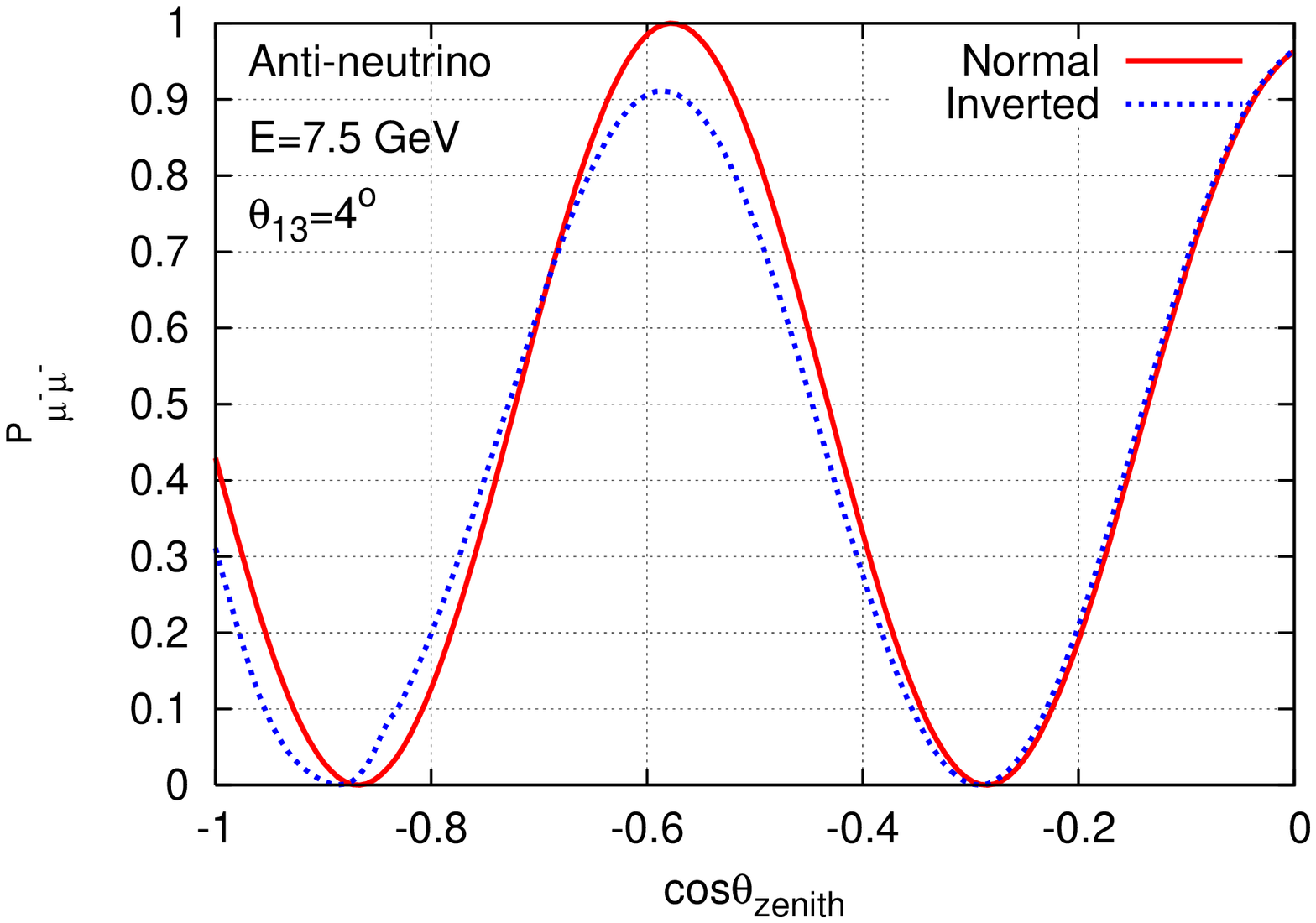}
\includegraphics[width=5.0cm,angle=270]{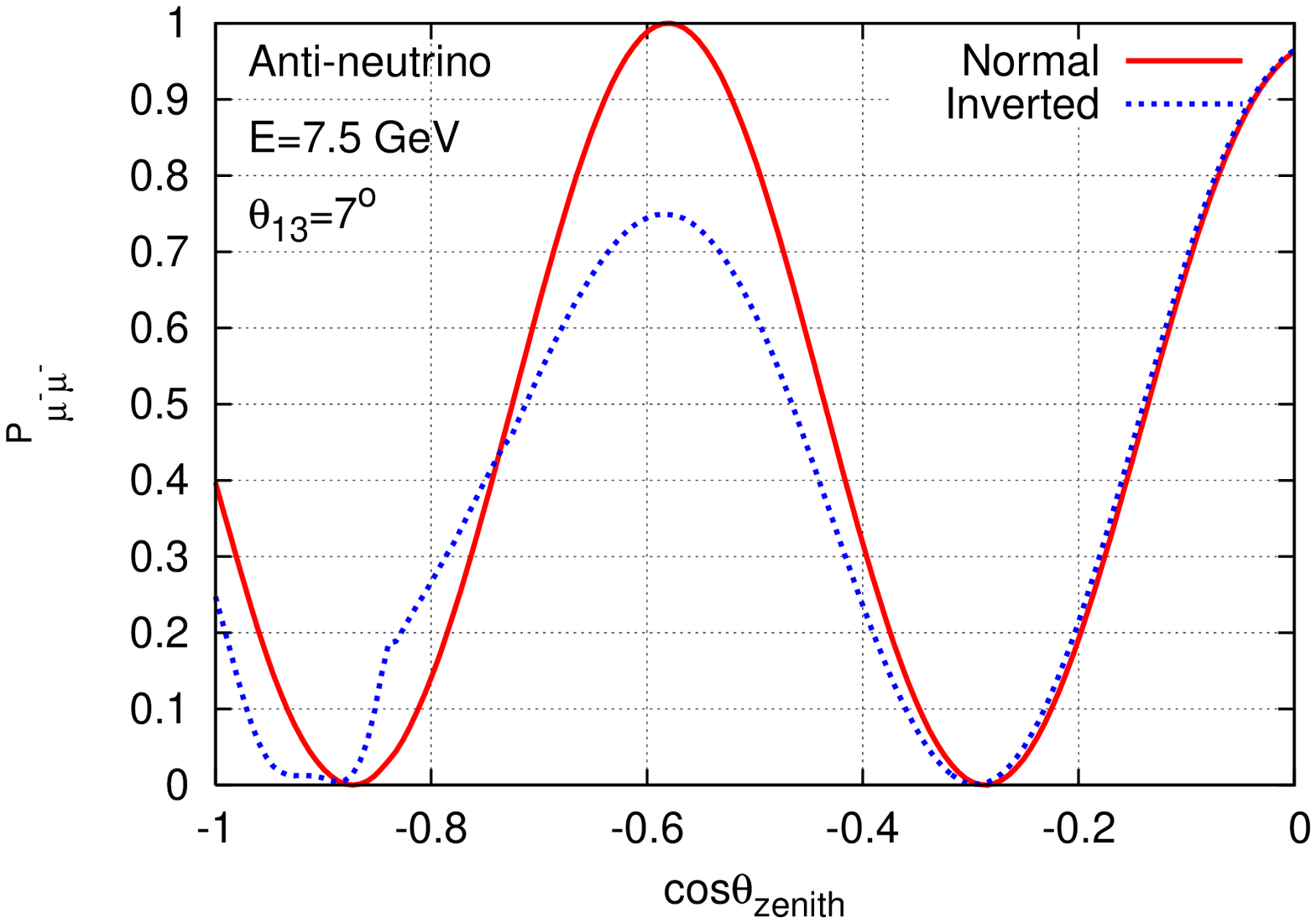}
\includegraphics[width=5.0cm,angle=270]{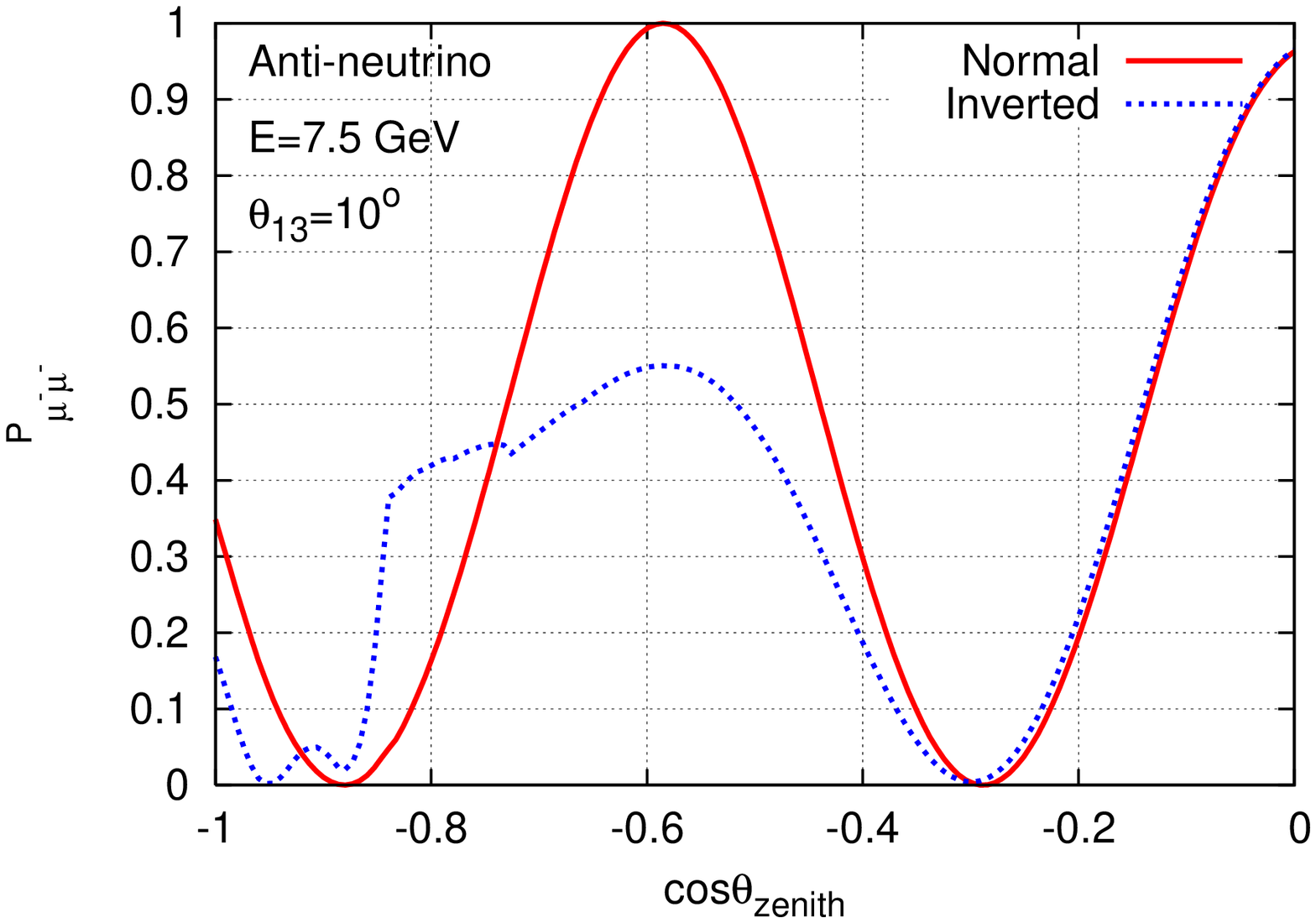}
\caption{\sf
The same plots of fig. \ref{f:p_nue3.7}, but with $E_\nu = 7.5$ GeV. 
}
\label{f:p_nue7.5}
\end{figure*}

\begin{figure*}[htb]
\includegraphics[width=5.0cm,angle=270]{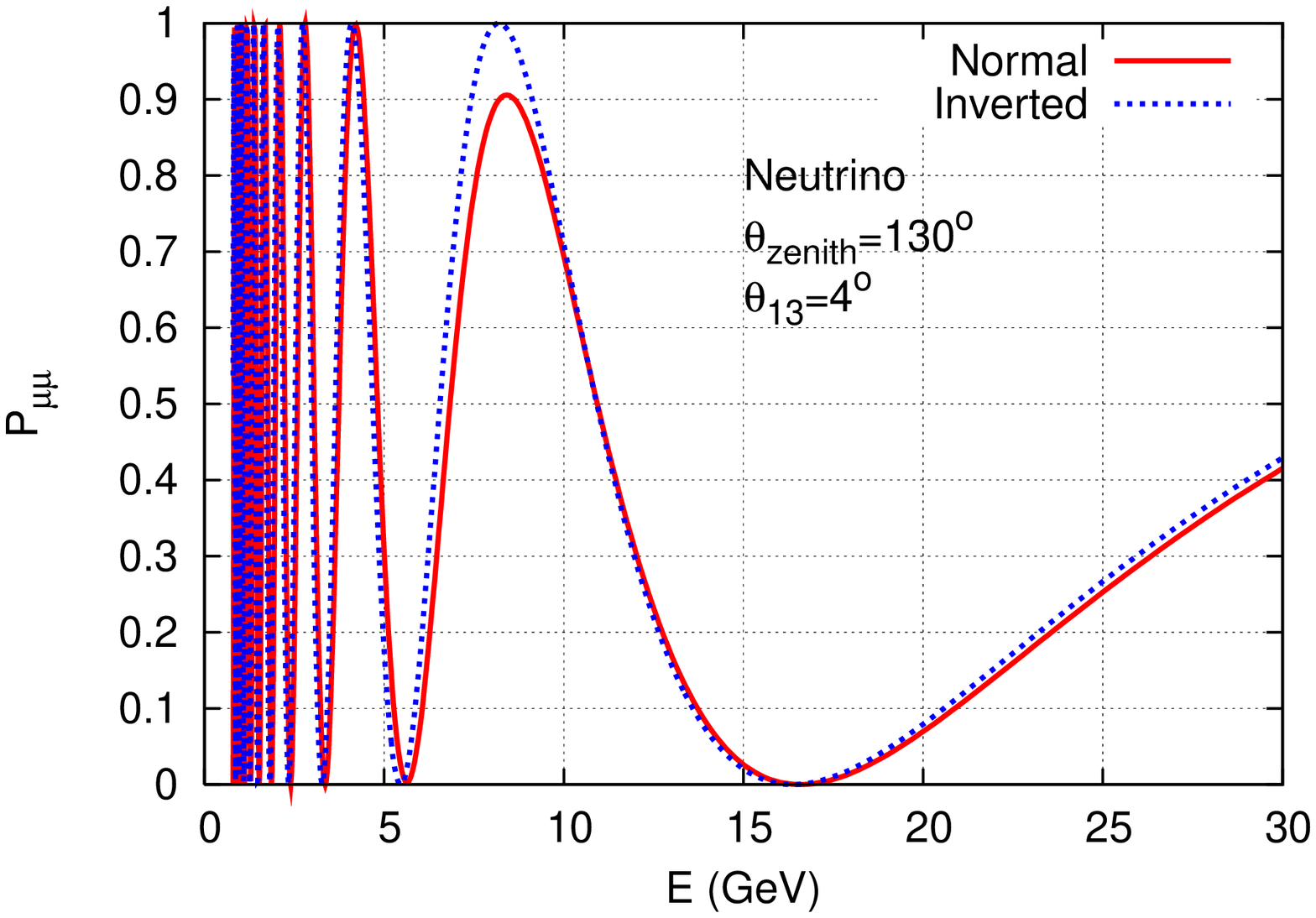}
\includegraphics[width=5.0cm,angle=270]{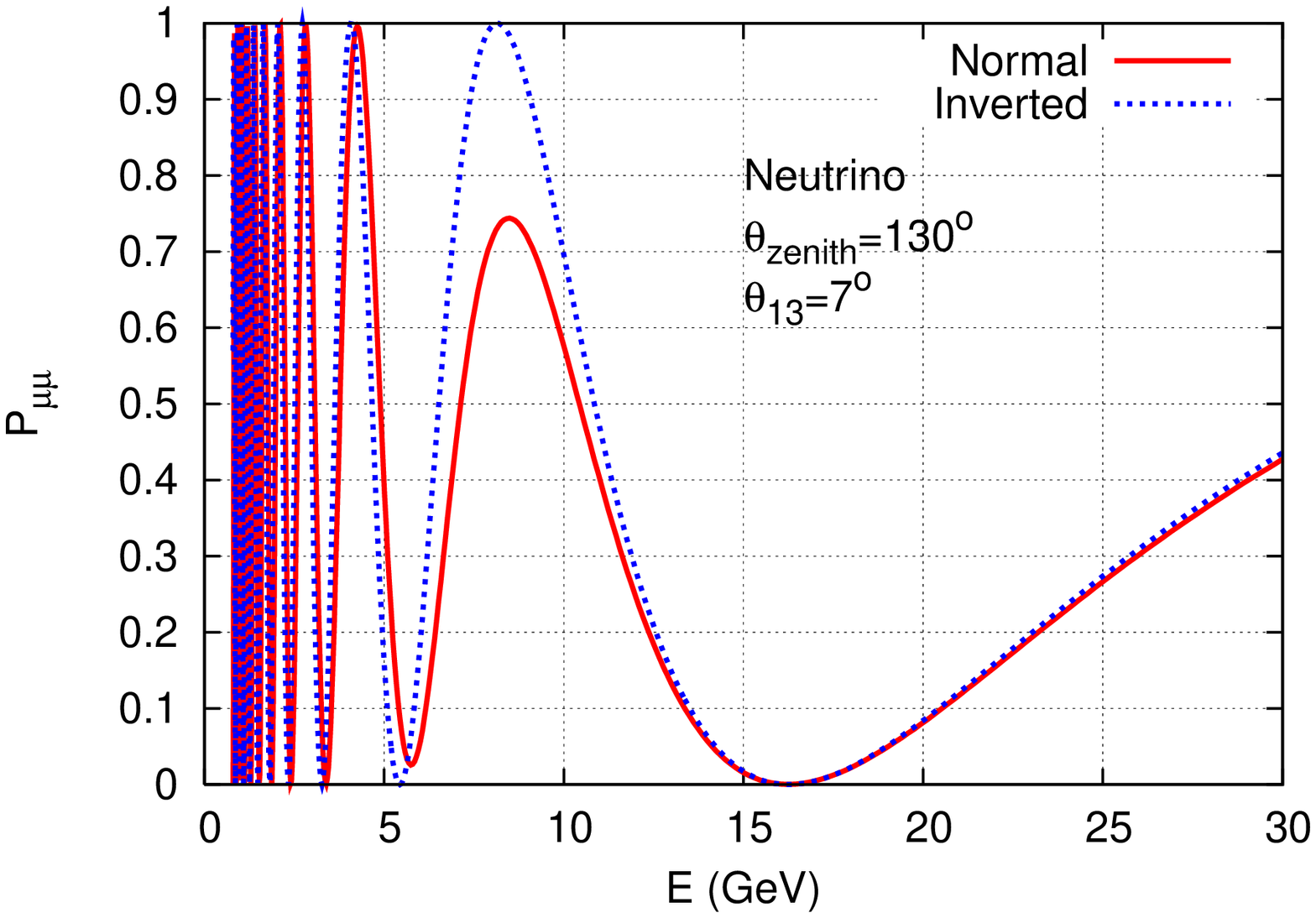}
\includegraphics[width=5.0cm,angle=270]{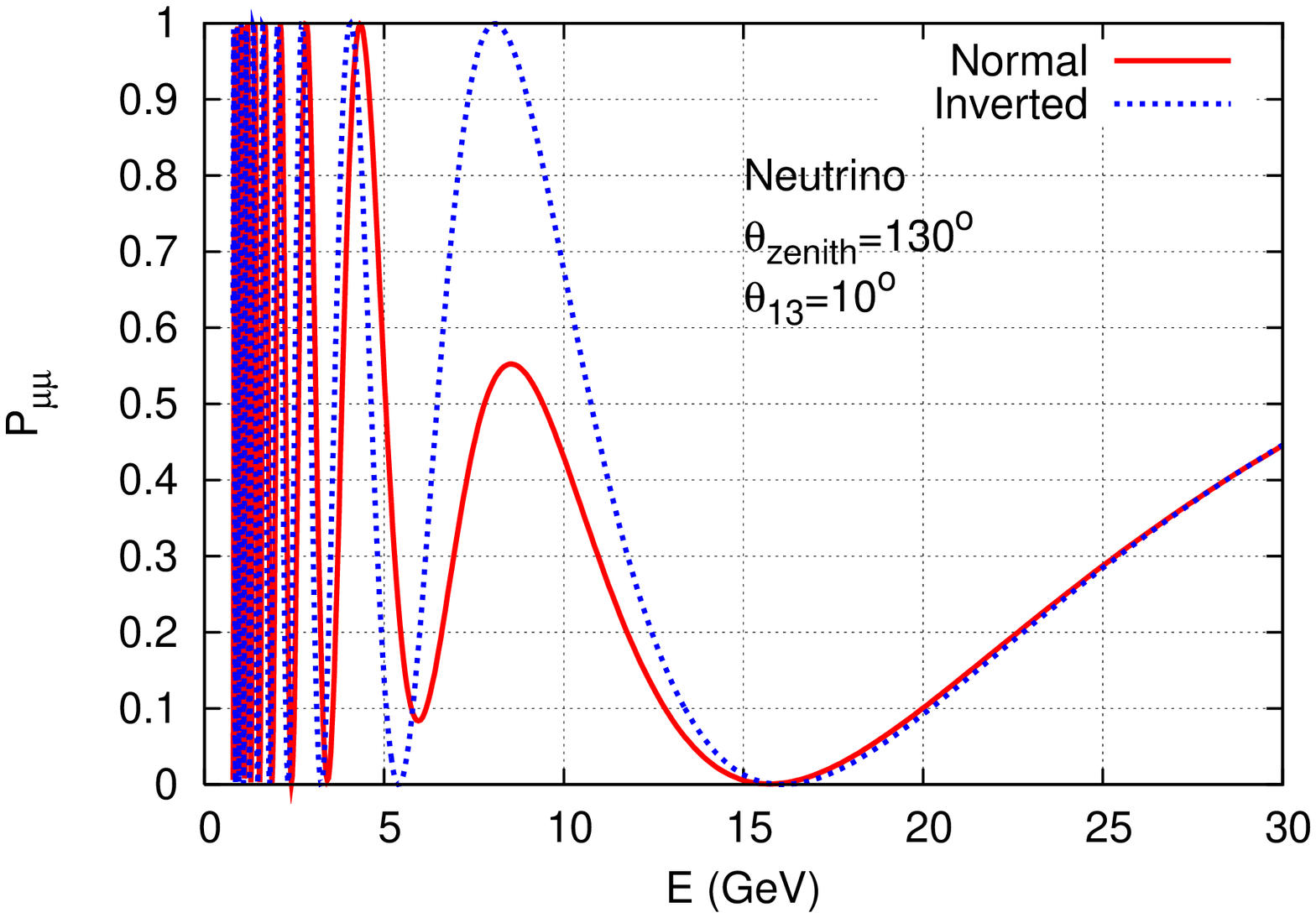}
\includegraphics[width=5.0cm,angle=270]{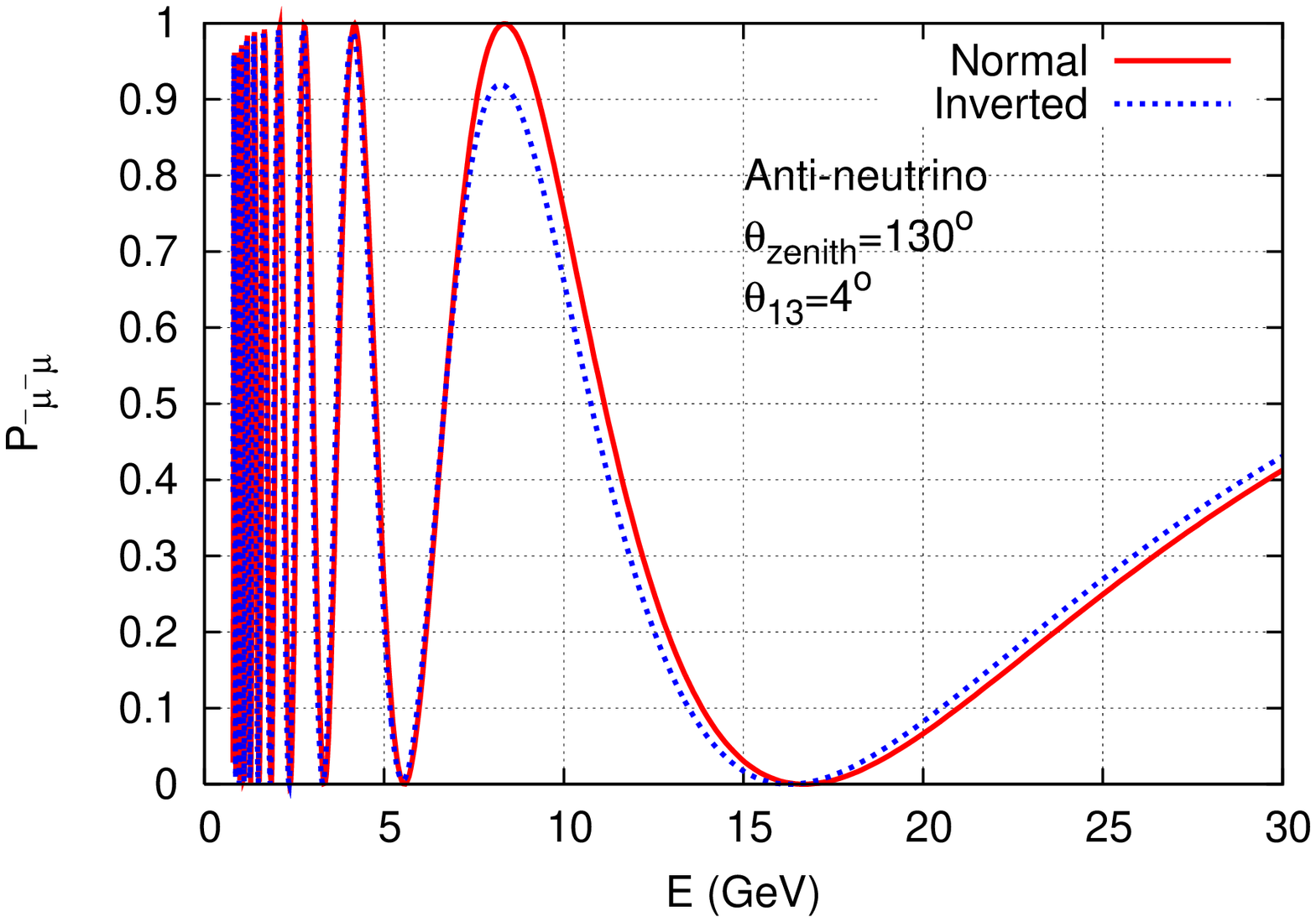}
\includegraphics[width=5.0cm,angle=270]{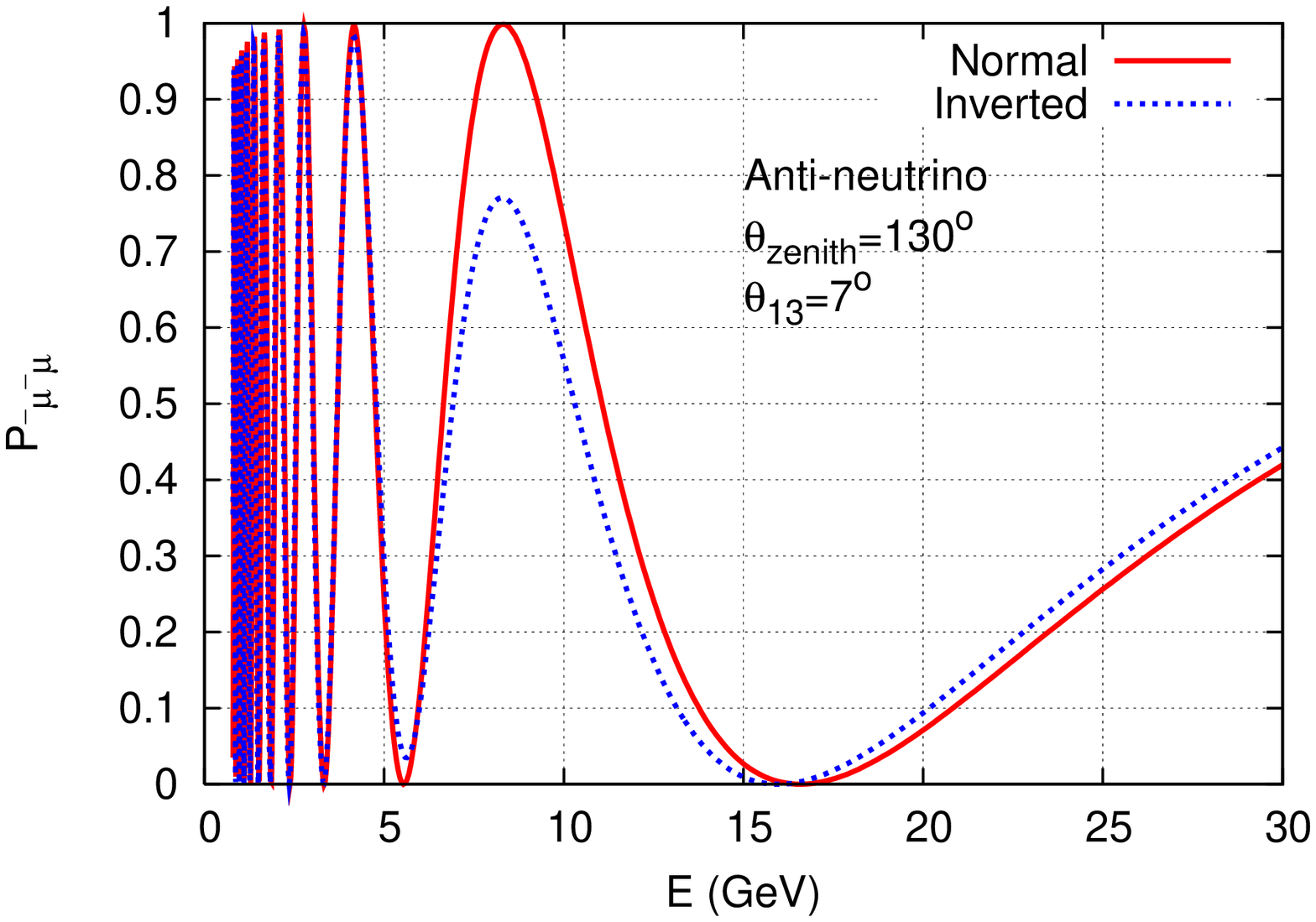}
\includegraphics[width=5.0cm,angle=270]{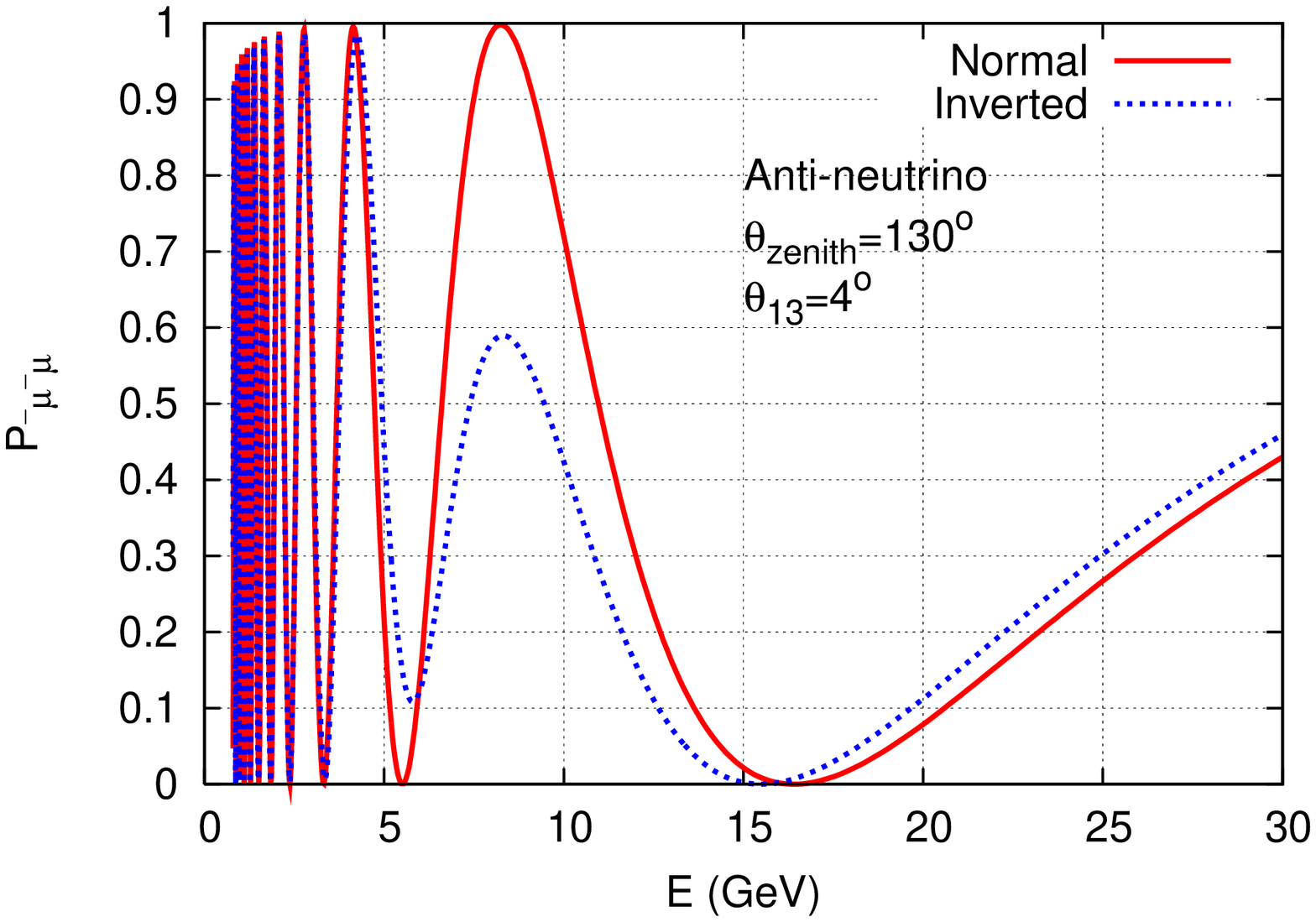}
\caption{\sf 
The variation of the $\nu_\mu$ (first row) and $\bar\nu_\mu$ (second row) 
survival probability with energy for $\theta_{13}=4^\circ$ (first column),
$7^\circ$ (second column), and $10^\circ$ (third column)  with a
typical zenith angle $\theta_{\rm zenith }=130^\circ$. }
\label{f:p_nuz}
\end{figure*}

Again, 
there is also a large depletion in survival probability of neutrino 
(anti-neutrino) for NH (IH) with respect to IH (NH) at the high energy ($\approx 5-10$ GeV) 
for  $-0.75 \gapp \cosz_\nu\gapp -0.40$
as seen in 
fig. \ref{f:p_nue7.5},    and \ref{f:p_nuz}. 

This pattern remains almost same over the present allowed range of 
oscillation parameters. 

Though all the above effects diminish rapidly as $\theta_{13}$ goes to zero,
there remains a finite difference in the survival probabilities
for NH and IH at $\theta_{13}=0$ due to nonzero value of 
$\Delta m_{21}^2$. 

\section{Possible systematic uncertainties}\label{s:uncertainties}

The systematic uncertainties 
may enter into the  
analysis of data 
in three steps:
i) flux estimation,
ii) neutrino interaction, 
and
iii) event reconstruction.

We can divide each of them into two categories: I) overall uncertainties (which are 
flat with respect to energy and zenith angle), and 
II) tilt  uncertainties (which are  function of  energy and/or zenith angle).

The  uncertainty in flux of category II may arise  due to the tilt in its shape 
with
energy and zenith angle.
This  can be expressed  in the following fashion for the case of energy:

\begin{equation}
      \Phi_{\delta_E}(E) = \Phi_0(E) \left( \frac{E}{E_0} \right)^{\delta_E}
      \approx \Phi_0(E) \left[ 1 + \delta_E \log_{10} \frac{E}{E_0} \right]
\label{e:uncer}
\end{equation}
This arises due to the uncertainty in spectral indices.
The uncertainty  ${\delta_E}=$5\% and $E_0 = 2$~GeV 
is considered in analogy with ~\cite{Kameda,skthesis2}. 

Similarly the flux uncertainty as a function of  zenith angle  can be expressed as 
\begin{equation}
      \Phi_{\delta_z}(\cos\theta_z) 
      \approx \Phi_0(\cos\theta_z) \left[ 1 + \delta_z \cos\theta_z \right]
\end{equation}
However,  this uncertainty is less  $(\approx 2\%)$ 
\cite{Kameda,skthesis2} than the energy dependent uncertainty.

{ The experimental systematic uncertainties may come through reconstruction 
of events.
Till now it is not estimated by the simulation of ICAL at INO. However the fact
from simulation is that the wrong charge identification possibility is
almost zero for the considered energy and zenith angle ranges. Since the detector
is symmetric to up and down going events, it is expected that the up/down
ratio cancels most of the experimental uncertainties.
}
The quantitative details of the uncertainties of category I (except those arising 
from the event reconstruction) 
can be found in \cite{Ashie:2005ik}.

\section{{ Migration from neutrino to muon energy and zenith angle}}\label{s:resolution}

In the CC processes, a lepton of same leptonic family of neutrino is produced 
with addition of some hadrons. Among the leptons, only  muon
produces a very clean track in ICAL detector and the hadrons produce a few hits
around the vertex of the interaction. Our analysis considers only muons,
and no hadrons for simplicity. 

For a given neutrino energy, the scattering angle and the
energy of the muon are related to each other. Again, the neutrino direction is 
specified by two angles: the polar angle and the azimuthal angle. 
Therefore, the actual resolution function
should be a function of energy and these two angles.

\begin{figure*}[htb]
\includegraphics[width=6.0cm,angle=270]{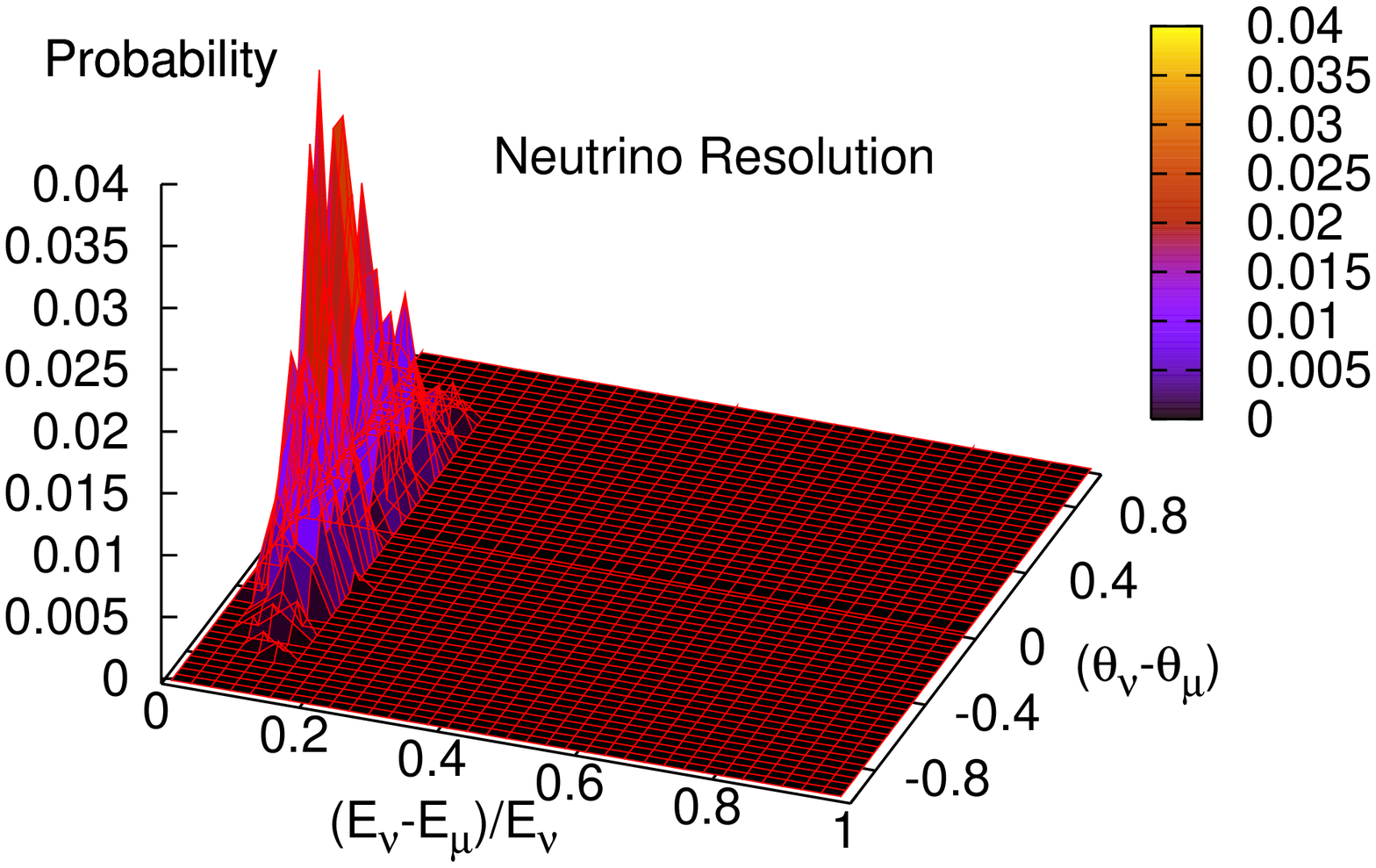}
\includegraphics[width=6.0cm,angle=270]{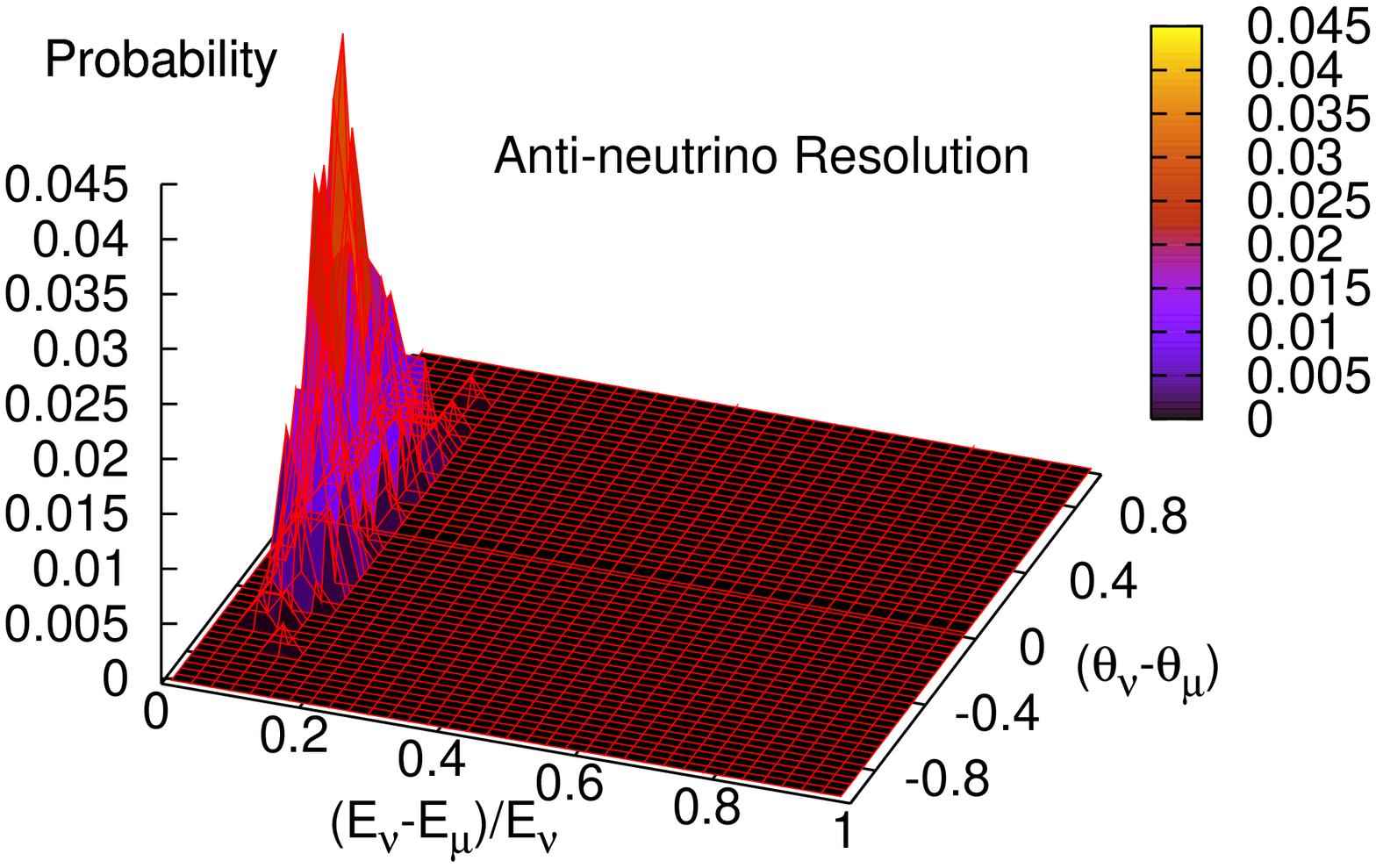}
\includegraphics[width=6.0cm,angle=270]{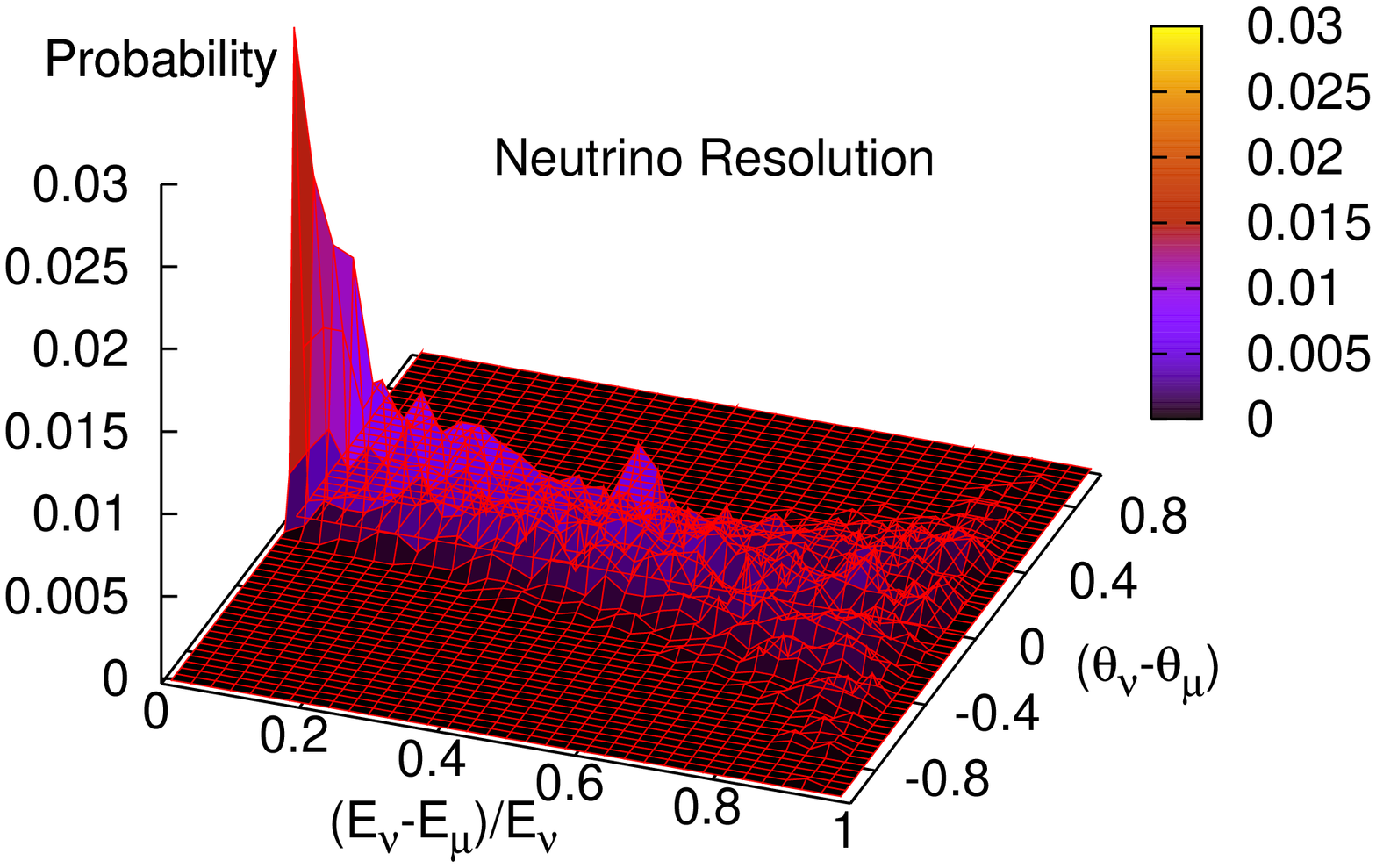}
\includegraphics[width=6.0cm,angle=270]{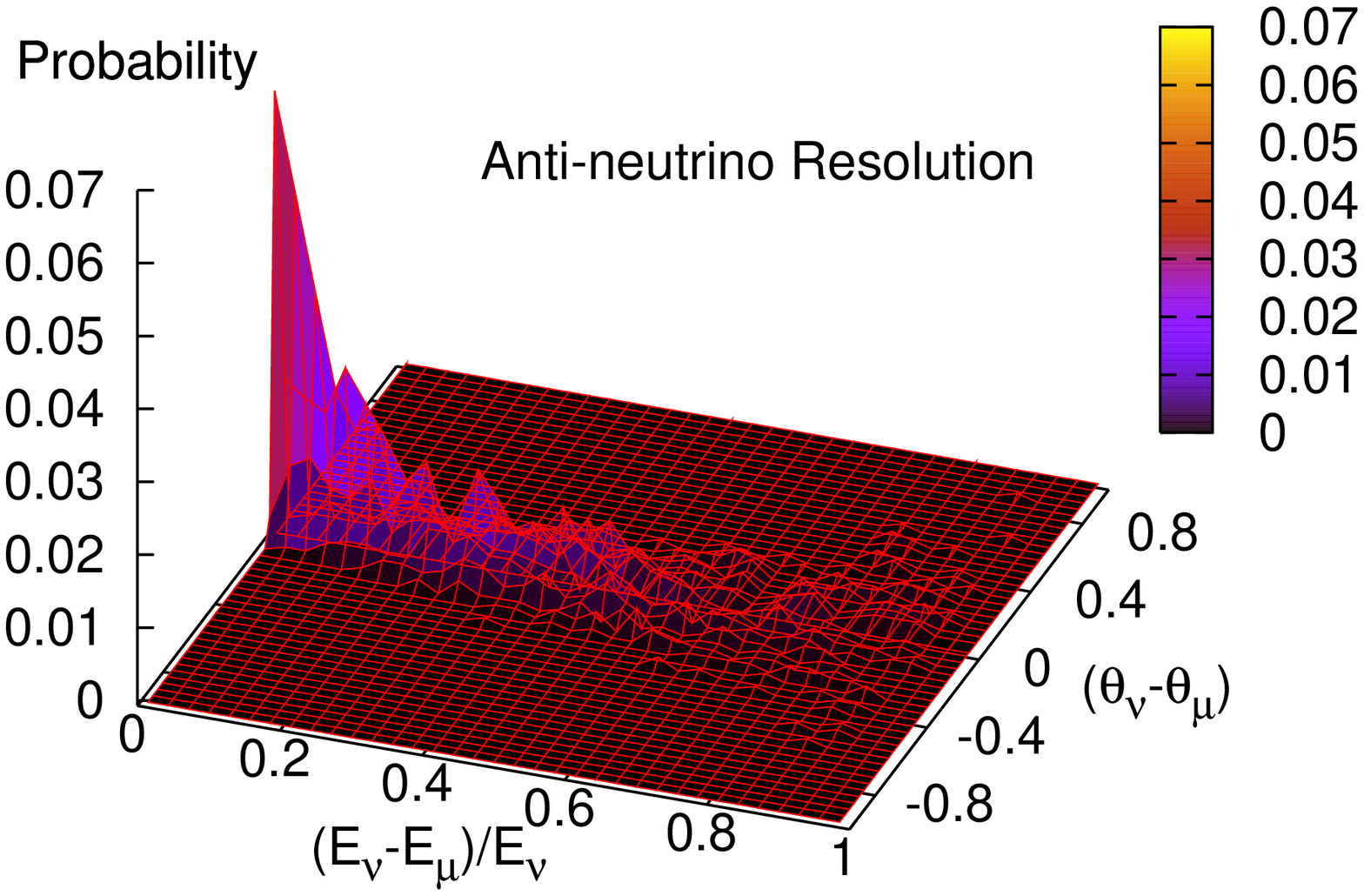}
\caption{\sf The distribution of probability of the resolution function
for  neutrino (left) and anti-neutrino (right)  at low energy ($\approx 1$ GeV) 
(top) and  at high
energy ($\approx 15$ GeV) (bottom).}
\label{f:resolution}
\end{figure*}

We are only interested in zenith angle which is a complicated function 
of above 
two angles. We reduce the above complexity 
of two angles by constructing a two dimensional resolution function
$R (x_{\rm E}, x_{\rm zenith})$ for a given neutrino energy with 
\be x_{\rm E}= \frac{( E_\nu - E_\mu)}{E_\nu}, ~~~~ 
x_{\rm zenith}=(\theta_\nu^{\rm zenith}-\theta_\mu^{\rm zenith}). \ee
To find the resolution function $R (x_E, x_{\rm zenith})$, we generate 
one thousand years un-oscillated atmospheric neutrino data for ICAL. 
Here, the two angles automatically come in proper way in zenith angle 
resolution and it should work well in analysis of the atmospheric data. 
We divide the whole data set into 17 neutrino energy bins (for $E=$0.8-18 GeV)
in logarithmic scale.  
It is also notable that the zenith angle resolution slightly depends on the 
value of neutrino zenith angle due to the spherical shape of the Earth.
To account for this dependence, we divide the whole zenith angle range 
($\cos\theta_z= -1~{\rm to}~ +1$) into 10
bins for every energy bin.
In fig. \ref{f:resolution} we show typical resolution functions at low ($\approx 1$ GeV) 
and high ($\approx 15$ GeV)
energy for both neutrinos and anti-neutrinos. 
Here, we see, 
the zenith angle resolution improves and
energy resolution worsens
with increase of energy. 
The zenith angle resolution is significantly
better for $\bar\nu_\mu$s than  $\nu_\mu$s.

From the distribution of resolution function (fig. \ref{f:resolution}), 
it is clear that the probability is 
high near zero of  $x_{\rm zenith}$ and it rapidly falls as we go 
far from it. Therefore we use a varying bin size in our analysis for both variables, small near 
zero and gradually bigger at far values.





{ How we  obtained the number of events in muon energy
and zenith angle bins from atmospheric neutrino flux using the above
resolution function 
is described in the following steps.

\begin{enumerate}
\item  For a given neutrino energy and zenith angle bin, we obtain
a distribution of events in a plane with axes $x_E=(E_\nu-E_\mu)/E_\nu$ and
$x_{\rm zenith}=(\theta_\nu^{\rm zenith}-\theta_\mu^{\rm zenith})$
from a large number of events generated by Nuance as shown in fig. 6.
From this distribution we find the probability of getting a muon at a
given energy and zenith angle bin
 dividing the  number of events of this bin by the total
number of events generated for the particular neutrino energy and zenith angle.

\item  The number of events for a given neutrino energy and zenith angle bin
is calculated by multiplying the exposed flux  with  the cross section
of this  energy. (The cross section is obtained from Nuance
as shown in Fig. \ref{f:cross}.)

\item  To obtain how many  events at a particular muon
energy and zenith angle bin may come  from the number of events of a given
neutrino energy and zenith angle bin,
we multiply this previously calculated number of events for the given neutrino
energy
with the probability  at this muon energy and angle bin obtained in step 1.

\item The total contribution at a particular muon energy and zenith angle
bin is obtained by adding the contribution from all neutrino energy and
zenith angle bins.
Finally we obtained the total muon events at different energy and
zenith angle bins in the considered ranges.
\end{enumerate}
}

\section{Choices of observables and its variables}
The effect of matter on the oscillation probability is a complicated 
function of $E$,   $L$ and the density of medium ($\rho$).
From the study of survival probability $P(\nu_\mu\leftrightarrow\nu_\mu)$ 
and  $P(\bar\nu_\mu\leftrightarrow\bar\nu_\mu)$ in section \ref{s:role}, 
it is clear that the difference between NH and IH 
is more significant for  particular ranges of $E$ and  $L$ rather than 
the range of the combination $L/E$. In case of the combination $L/E$, 
there remains
some $L(E)$ values for some values of $E(L)$ where the difference  between
NH and IH  is insignificant. For these values of $L$ and $E$, it increases 
the magnitude of statistical errors and eventually kills the significance 
which comes from the interesting regions of $L$ and $E$. For this reason, we need to consider only  important ranges of the baseline $L_i$ for each range of  
energy $E_i$.

We may consider the following observables to determine the hierarchy:
$u_i,~ d_i,~ \bar u_i,~ \bar d_i$,~ ${u_i}/{d_i}$,~ ${\bar u_i}/{\bar d_i}$, 
or any combinations of them, 
where, $ u_i,~ d_i~ (\bar u_i,~ \bar d_i)$ are the  up
going and down going neutrino (anti-neutrino) events in the 
interesting $L_i-E_i$ regions.
For down going events we consider the `mirror' $L$ which is exactly
equal to that value when the neutrino comes from the opposite direction. 

In case of observables with the combination of neutrinos and anti-neutrinos, 
we will encounter the uncertainty of the ratio ${\nu_\mu}/{\bar\nu_\mu}$ which is
$\simeq 1$ at low energy and it increases with  energy \cite{Honda:2004yz,
Battistoni:2003ju,Barr:2004br}.
Since the matter effect and 
the cross sections for neutrinos and anti-neutrinos are dependent on $E$, 
the uncertainty in the ${\nu_\mu}/{\bar\nu_\mu}$ ratio may affect 
significantly in determination of hierarchy for the combined  observables.

Firstly, 
we choose the observables as a ratio of $u$ and $d$
to { cancel  all the overall uncertainties} (discussed in section 
\ref{s:uncertainties}).  
Secondly, since the  magnetized 
ICAL detector has the capability to distinguish $\mu^-$ and $\mu^+$, we consider the 
ratios separately for $\nu_\mu$s and $\bar\nu_\mu$s:
\be A= { u}/{ d}, ~~~{\rm and}~~~\bar A= {\bar u}/{\bar d}.\ee

In fig. \ref{f:resonance},  it clearly shows that there is  a sudden fall in resonance
energy due to a sudden rise in average density of the Earth. We choose two ranges of 
energy: one for  core (vertical events)
and  another for mantle (slant events).  The difference in survival 
probability between NH and IH can also be seen in figs. 
 \ref{f:p_nue3.7}, and
\ref{f:p_nue7.5}
at these 
two ranges of zenith angle for two typical resonance energy of 3.7 GeV and 7.5 GeV, respectively.
The resonance range of energy for a typical value of the zenith angle  is also shown
in fig. \ref{f:p_nuz}.  

From the discussion of section \ref{s:flux}, \ref{s:role} and 
\ref{s:resolution}, we observed the following : 
\begin{enumerate}
\item The zenith angle resolution width 
is  large
with a peak around zero
(fig. \ref{f:resolution}).

\item
From the distribution of the energy resolution, it is clear that
a muon of energy $E_\mu$ can come from any neutrino having energy 
$E_\nu \gapp E_\mu$
with almost equal probability for $E_\nu \gapp$ 3 GeV where the deep inelastic
events dominate.
%
Again, as the flux falls very rapidly with energy (see fig. \ref{f:cross})
and  $E_\nu \gapp E_\mu$ (this is obvious for $E_\nu > 1$ GeV for targets at
rest), the lower value of the energy range will play a crucial role in the results.

\end{enumerate}

The bonus point of these two  ranges of zenith angle (vertical and slant) 
is that the vertical/horizontal 
uncertainty in flux is minimized and becomes negligible. 
Among the systematic uncertainties discussed in section 
\ref{s:uncertainties}, we remain with the energy dependent flux uncertainty. 
Here we will not consider any uncertainty coming from reconstruction of the 
events.
This will be taken care of in near future in GEANT-based studies.

\section{$\chi^2$ analysis}
Here we are interested to find the hierarchy discrimination
ability of ICAL@INO at low $\theta_{13}$ using the resonance ranges. 
It should be noted here that the resonance
width falls very rapidly with $\theta_{13}$.

From the earlier works \cite{Petcov:2005rv,
Gandhi:2007td,Indumathi:2004kd}   it is found that as the energy and the 
angular resolution width increase the discrimination
ability of hierarchy becomes  worsened (see fig. 10 of \cite{Gandhi:2007td}). 
In our actual simulation it is found that the width of these resolutions 
are large (see fig. 6).

We have tried to tackle the problem in the following way.
We do not  divide the selected ranges. We calculate the
total up-going ($u_i$)  and the total down-going ($d_i$) events and 
then their ratio ($u_i/d_i$) for each selected range.
Finally we  perform the chi-square study with these $u_i/d_i$ ratios of
all ranges (see table \ref{t:range}).

\begin{figure*}[htb]
\includegraphics[width=5.0cm,angle=270]{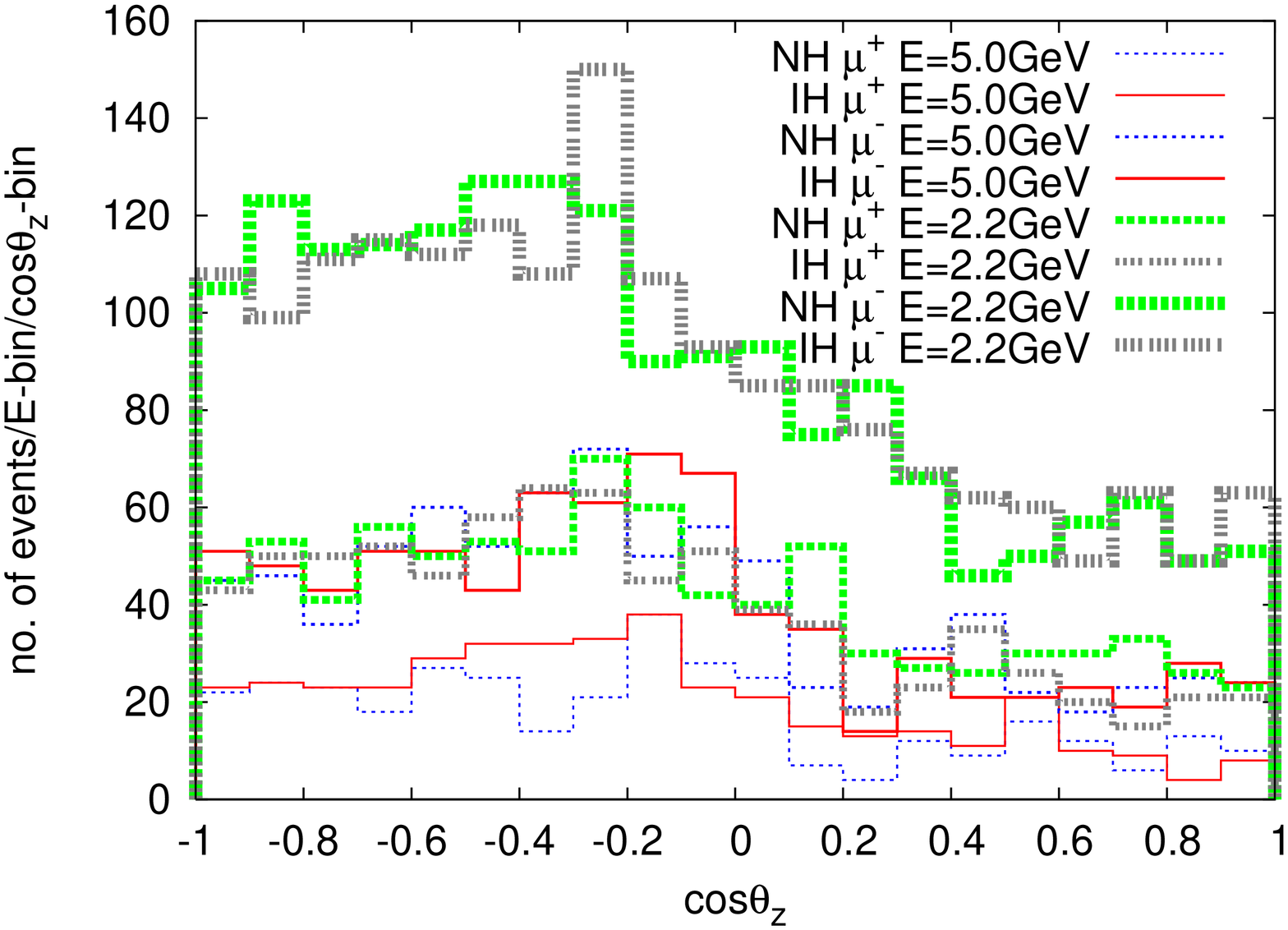}
\includegraphics[width=5.0cm,angle=270]{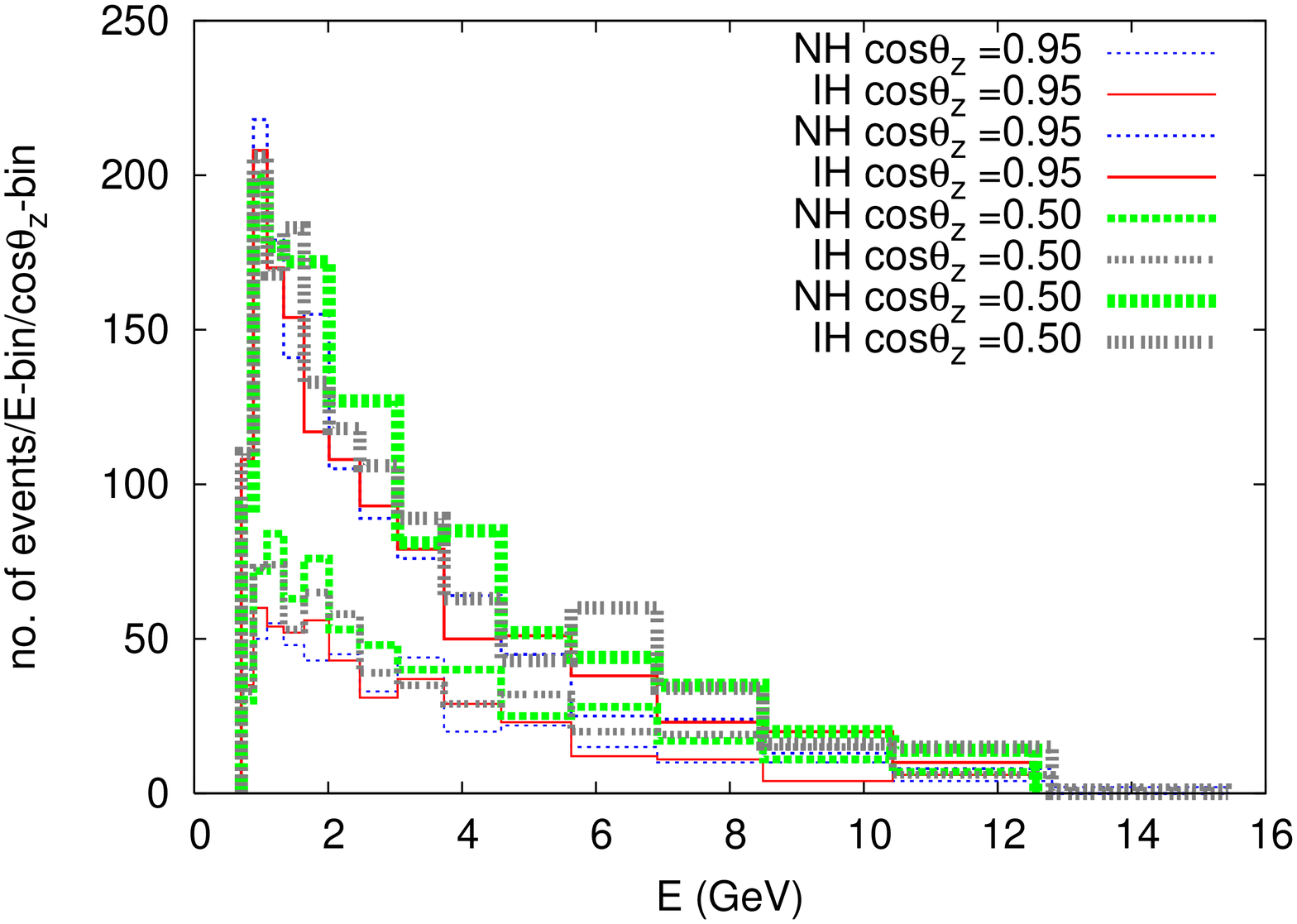}
\caption{\sf  The typical distribution of the number of events with 
$\cos\theta_{zenith}^\mu$ for fixed $E_\mu$ (left) 
and with $E_\mu$ for fixed $\cos\theta_{zenith}^\mu$ (right).
 We set $\Delta m_{32}^2=-2.5\times 10^{-3}$eV$^2$,
 $\theta_{23}=45^\circ$ and $\theta_{13}=7.5^\circ$.}
\label{f:events}
\end{figure*}

For a set of input values of the oscillation parameters,
we generate the neutrino events for 1 MTon.year exposure of ICAL detector using the Nuance
(see fig. \ref{f:events}). 
Next we calculate 
the total $u_i$ and the total $d_i$  
for each considered range of 
$E_{\mu_i}-\cosz_{\mu_i}$ (see table \ref{t:range}) 
and then their  ratio $u_i/d_i$. 
We call it the ``experimental data".
The statistical error is estimated by the following 
formula
\be \sigma_{u_i/d_i} = \sqrt{\frac{1}{u_i}+\frac{1}{d_i}}
\left (\frac{u_i}{d_i}\right ).
\label{e:staterr}\ee

{  We adopt the pull method for chi-square analysis. 
The theoretical data in chi-square  is calculated with 
the following numerical method.
\begin{enumerate}
\item
Considering a set of the oscillation parameters 
and a value of the pull variable for a considered systematic uncertainty
(using eq. \ref{e:uncer}), we obtain the neutrino flux in 200 bins of 
$E_\nu$ and 300 bins of $\cosz_\nu$. At low energy region the number of oscillation 
swings with $E$ ($L$) for a fixed $L$
($E$) are very large (see fig. \ref{f:p_nue3.7} and \ref{f:p_nuz}). Again the difference of the oscillation probabilities
between NH and IH changes very rapidly for some ranges of $L$ and $E$.
To obtain the accurate oscillation pattern one needs finer binning of
$L$ and $E$. We have checked the oscillation plots varying the number
of bins for $E_\nu$ and $\cos\theta^{\rm zenith}_\nu$ and finally fix
200 bins for $E_\nu$ and 300 bins for $\cos\theta_\nu^{\rm zenith}$.
We have used same number of bins to obtain both experimental and theoretical 
data. 
The number of events from neutrino to muon energy and zenith angle bins are 
obtained using the resolution function and the cross section as discussed 
earlier.
\item
Next we calculate  the total $u_i$ and the total 
$d_i$  separately for each selected range of 
$E_{\mu_i}$-$\cosz_{\mu_i}$ 
(see table \ref{t:range}).
Then  the ratio  $u_i/d_i$ is determined for each  range.
We call it the ``theoretical data".  It should be noted that 
the number of bins has been  changed in chi-square from that of the flux.
(The reason is described above.)

\item
 We calculate the standard chi-square from these ``theoretical
 data", the above ``experimental data"  and the ``statistical error"
(eq. \ref{e:staterr}) for all ranges.  
Next the minimization of this chi-sqaure is done with respect to the 
pull variable to find its best-fit value keeping all other 
oscillation parameters fixed.
Finally the  ``theoretical data" is again calculated
using the  best-fit value of the pull variable. 
\end{enumerate}

In the last the $\chi^2$  is calculated from the final ``theoretical data"  and the 
above ``experimental data".  The whole process is done
to find the $\chi^2_\mu$ and $\chi^2_{\bar\mu}$   
for $\nu_\mu$s  and $\bar\nu_\mu$s, respectively. 
At the end, we find the total
$\chi^2=\chi^2_\mu + \chi^2_{\bar\mu}$ for the considered set of oscillation
parameters. 

As discussed in section \ref{s:uncertainties}, we consider only neutrino
flux uncertainty due to its tilt with energy\footnote{
It should be noted that
for the conservativeness of our results
we do not consider any contribution to $\chi^2$ from the 
``prior" of the oscillation parameters 
that may come from other experiments.}.

\section{ Resonance ranges in terms of $E_\mu$ and $\theta_\mu^{\rm zenith}$}
The optimization of the ranges  by analytical calculation is almost impossible for the
complexity in the behavior of survival probability,
the smearing of resolution, and the rapid fall of flux with increase of energy.

We calculate the $\chi^2$ varying the energy and zenith angle ranges
for both NH and IH for a given input of $\theta_{13}$ and type of hierarchy
keeping all other  oscillation parameters at  
 their best-fit values. 
Then, we select 
the ranges where the  $\chi^2$ difference for NH and IH is the largest.
This is done for the input values of $\theta_{13}=4^\circ, 5^\circ, 7^\circ, 10^\circ$
and tabulated in table \ref{t:range}. Due to the effect of energy resolution
function and the nature of flux, the higher end of the energy range can not
be precisely determined. We see, the ranges squeeze rapidly with the decrease
of $\theta_{13}$. It should be noted that the ranges determined
here has a significant  weight of
the atmospheric neutrino flux and it is not independent of flux since
it is not from the migration of neutrinos to muons only.
However, there is no significant change in chi-square if
the ranges are changed within few percent of their boundary values.
\begin{table*}[htb]
\begin{center}
\begin{tabular}{|c|c|c|c|c|}
\hline
\multicolumn{5}{|c|}{Resonance ranges in terms of  muon energy 
and zenith angle 
 }\\
\hline
\multicolumn{1}{|c|}{$\theta_{13}$} & \multicolumn{2}{|c|}{Core}  & 
\multicolumn{2}{|c|}{Mantle}\\
\cline{2-5}
 & $\cos\theta^{\rm zenith}_\mu$ & $E_\mu$ (GeV)&
$\cos\theta^{\rm zenith}_\mu$ & $E_\mu$ (GeV)\\
& $\mu^-$ ~~~~~~~~$\mu^+$ &  $\mu^-$~~~~ ~~~~$\mu^+$ &  $\mu^-$~~~~~~~~ $\mu^+$ & $\mu^-$~~~~~~~~$\mu^+$\\

\hline 
$10^\circ$ &-1.0 to -0.57~~~~-1.0 to -0.72&0.79 to 9.40~~~~0.93 to 7.40& -0.57 to -0.41~~~~-0.68 to -0.46 & 3.33 to 8.01~~~~5.38 to 8.68\\
$7^\circ$ &-1.0 to -0.63~~~~-1.0 to -0.79& 1.39 to 10.18~~~~2.07 to 4.59 & -0.61 to -0.43~~~~-0.70 to -0.46 & 4.24 to 6.83~~~~5.38 to 11.94 \\
$4^\circ$&-1.0 to -0.50~~~~-1.0 to -0.83  & 1.63 to 8.68~~~~2.07 to 6.83  & -0.48 to -0.30~~~~-0.72 to -0.46 &  4.59 to 6.31~~~~7.40 to 10.18 \\ 
\hline
\end{tabular}
\end{center}
\caption{\sf The resonance ranges of  $\cos\theta^{\rm zenith}_\mu-E_\mu$ 
for different input values of $\theta_{13}$. The type of input hierarchy is 
inverted.}
\label{t:range}
\end{table*}

\section{Marginalization and Results}

The neutrino oscillation parameters can not be measured at infinite
accuracies.  To find at what statistical significance the wrong
hierarchy can be disfavored, we calculate the $\chi^2$ for both true
and false hierarchy 
varying the oscillation parameters in the following ranges (unless 
we specify) :
\begin{enumerate}
\item $|\Delta m^2_{23}|$ :  $2.3\times 10^{-3} -2.8\times 10^{-3}$eV$^2$,
\item $\theta_{23}$ :  $39^\circ -51^\circ$, and
\item $\theta_{13}$ :  $0^\circ-12^\circ$.
\end{enumerate}
We set other oscillation parameters at their best-fit values and choose 
CP-violating phase $\delta=0$.
For the data set of each input values of $\theta_{13}$, 
we used the corresponding
resonance range of $E_\mu - \cosz_\mu$.

The marginalized $\chi^2$ { for both true and false hierarchy }as a function of $\theta_{13}$ is shown 
in fig. \ref{f:marginmt13} for the input of IH and different 
values of 
$\theta_{13}$ ($5^\circ, 7^\circ$ and $10^\circ$). 
Here the marginalization is done over other two oscillation
parameters $|\Delta m_{32}^2|$ and $\theta_{23}$. 
The similar plot is also shown for the input of NH in 
fig. \ref{f:marginpt13}, but with $E_\mu-\cosz$ ranges as optimized 
for IH.
{ Finally we show the discrimination sensitivity $\Delta \chi^2 = 
\chi^2 ({\rm false})-\chi^2 ({\rm true})$ 
marginalized over all three oscillation parameters
$|\Delta m_{32}^2|$, $\theta_{23}$ and $\theta_{13}$ }
as function of input values of $\theta_{13}$ in fig. \ref{f:margin}. 

The changes reflected in $\chi^2$ with the input of $\theta_{13}$ and the type of hierarchy 
can be understood from the following facts:
\begin{enumerate}
\item Normally the number of events is larger for $\nu_\mu$ than $\bar\nu_\mu$ 
due to larger $\nu_\mu$ cross section. So the $\nu_\mu$ contribution to the 
difference in 
chi-square due to hierarchy is larger than $\bar\nu_\mu$.  Since $\nu_\mu$s 
are suppressed in NH, the input with IH (larger statistics) 
will give a larger difference in 
chi-square and
NH can be more  easily disfavored
than IH. (See fig. \ref{f:marginmt13} 
and  \ref{f:marginpt13} for the input value of $\theta_{13}=10^\circ$.) 
However, once any one type of the hierarchies is disfavored, there remains 
the other
type only. So it is  essential to disfavor any one of them.

\item The resolution is better for anti-neutrinos than neutrinos.
 Again as  $\theta_{13}$ decreases, the 
resonance ranges  squeeze very rapidly. 
This leads to a competition between
resolution and statistics  resulting a comparable
contribution of $\nu_\mu$ and $\bar\nu_\mu$ to chi-square difference for $\theta_{13} \lapp 7^\circ$
(compare fig. \ref{f:marginmt13}
and  \ref{f:marginpt13} for the input value of $\theta_{13}=7^\circ$).


\item 
%

Though there is an appreciable difference in survival probabilities between
 NH and IH
 below $\theta_{13}$ value of $5^\circ$, 
the tilt  uncertainty with energy kills the difference in chi-squares.
\end{enumerate}
\begin{figure*}[htb]
\includegraphics[width=4.0cm,angle=270]{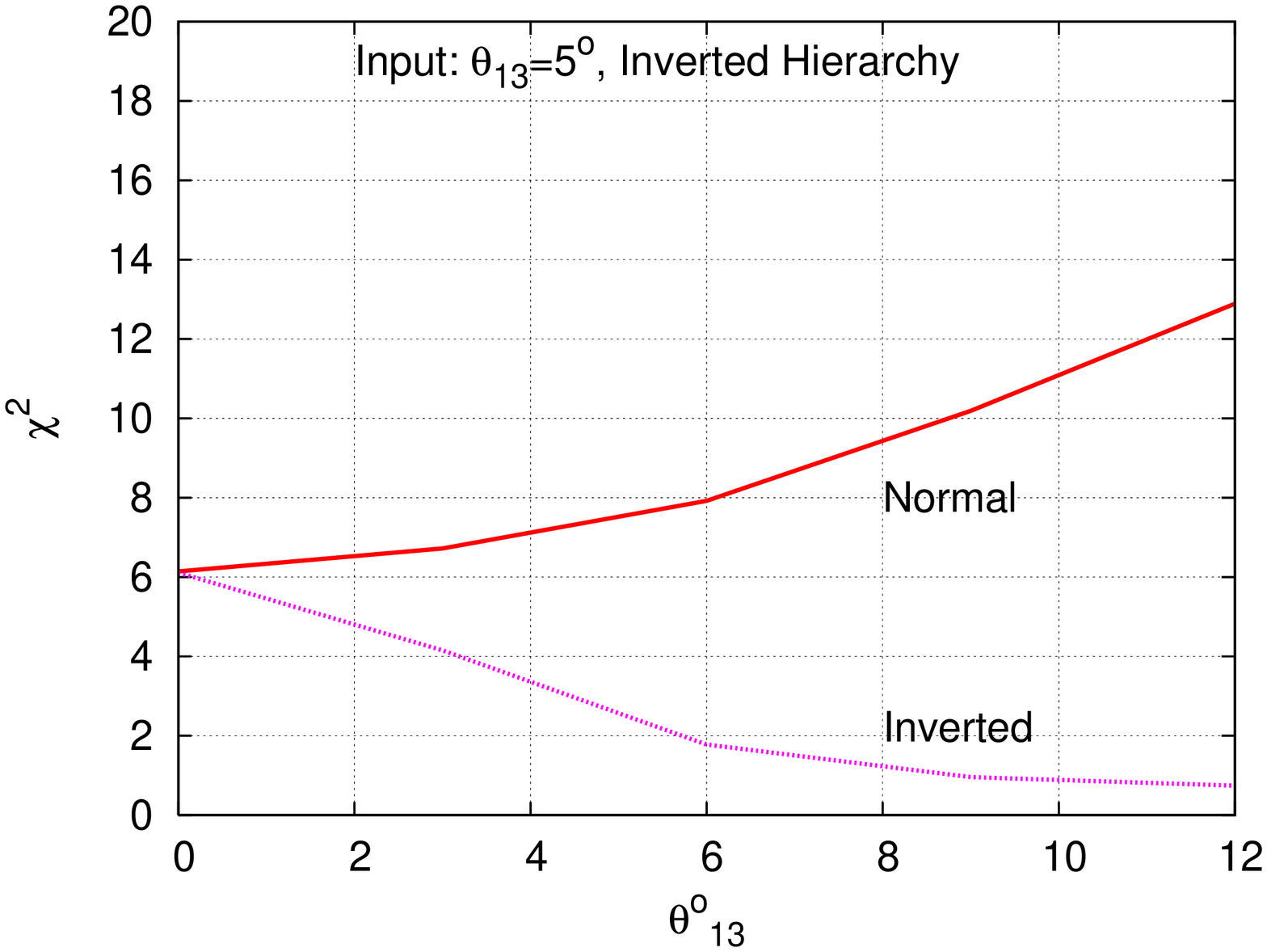}
\includegraphics[width=4.0cm,angle=270]{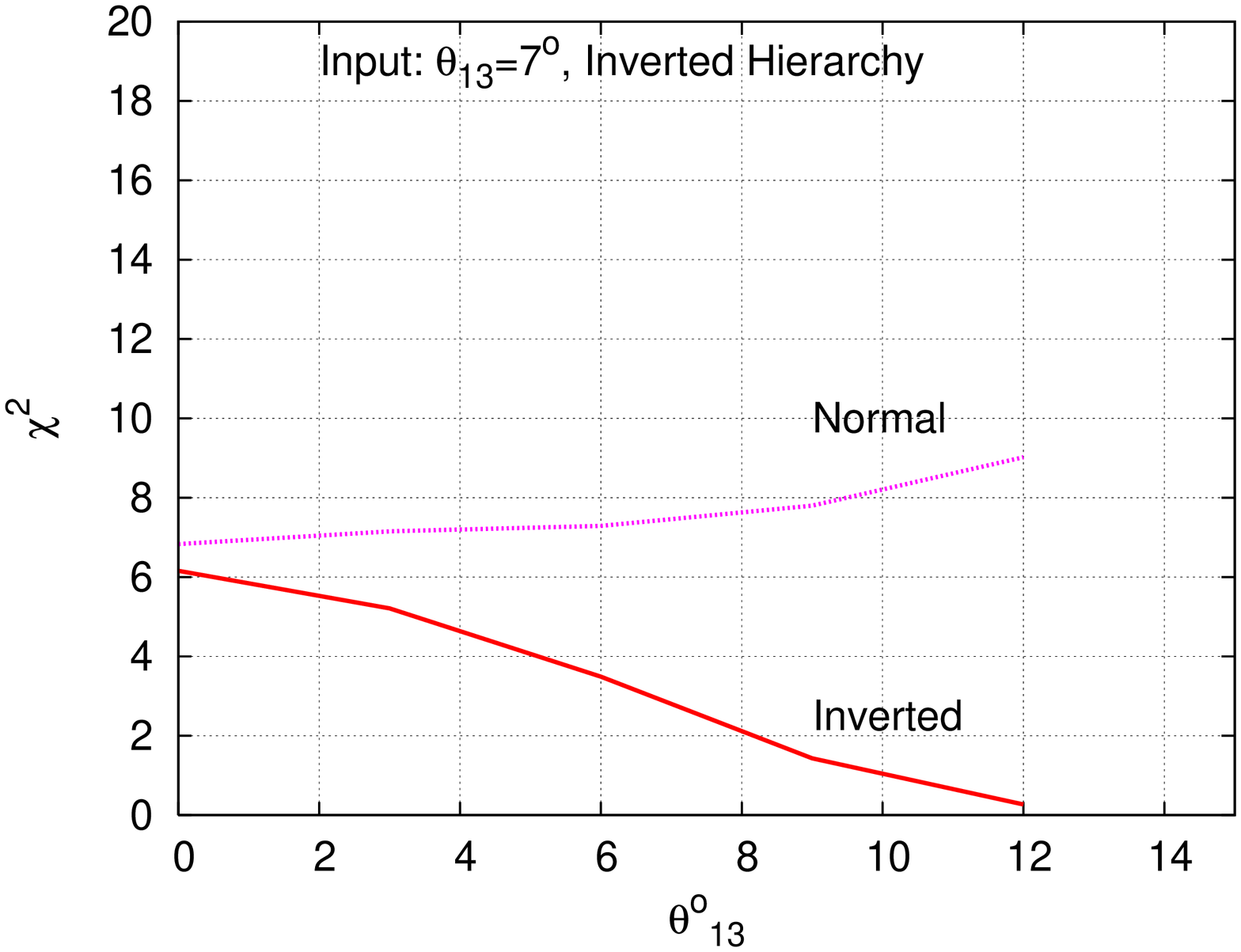}
\includegraphics[width=4.0cm,angle=270]{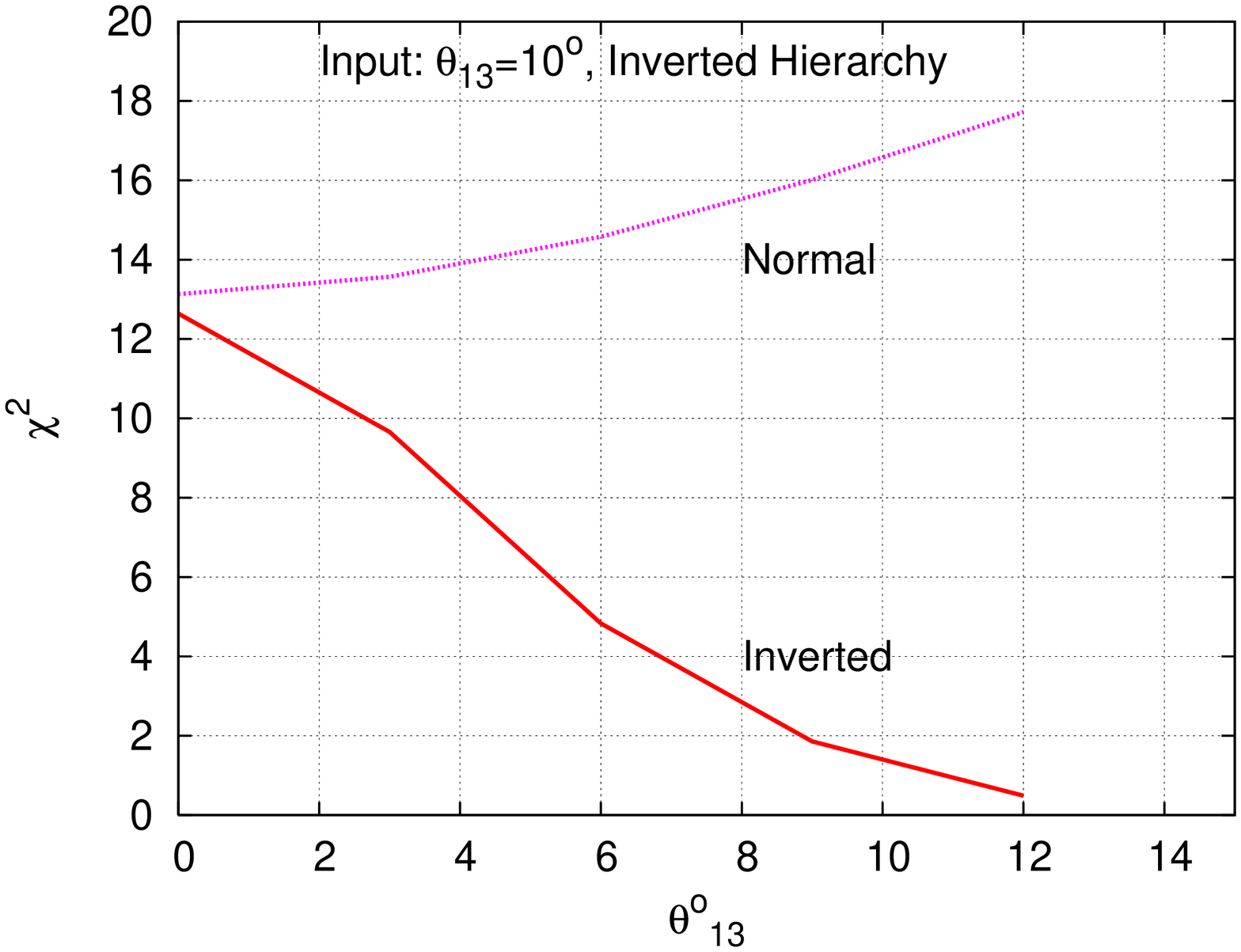}
\caption{\sf 
The variation of marginalized $\chi^2$  with $\theta_{13}$
for NH and IH for the input of  IH
and the  values of 
$\theta_{13}$=$5^\circ$ (left), $7^\circ$ (middle) $10^\circ$ (right).
The marginalization is done over $|\Delta m_{32}^2|=(2.3-2.7)\times 10^{-3}$eV$^2$
and $\theta_{23}=39^\circ-51^\circ$. 
We consider a uncertainty in tilt
factor $\delta_E=5\%$ at $E_\nu$=2 GeV. 
The other oscillation parameters are
set at their best-fit values (discussed in the text). 
}
\label{f:marginmt13}
\end{figure*}

\begin{figure*}[htb]
\includegraphics[width=5.0cm,angle=270]{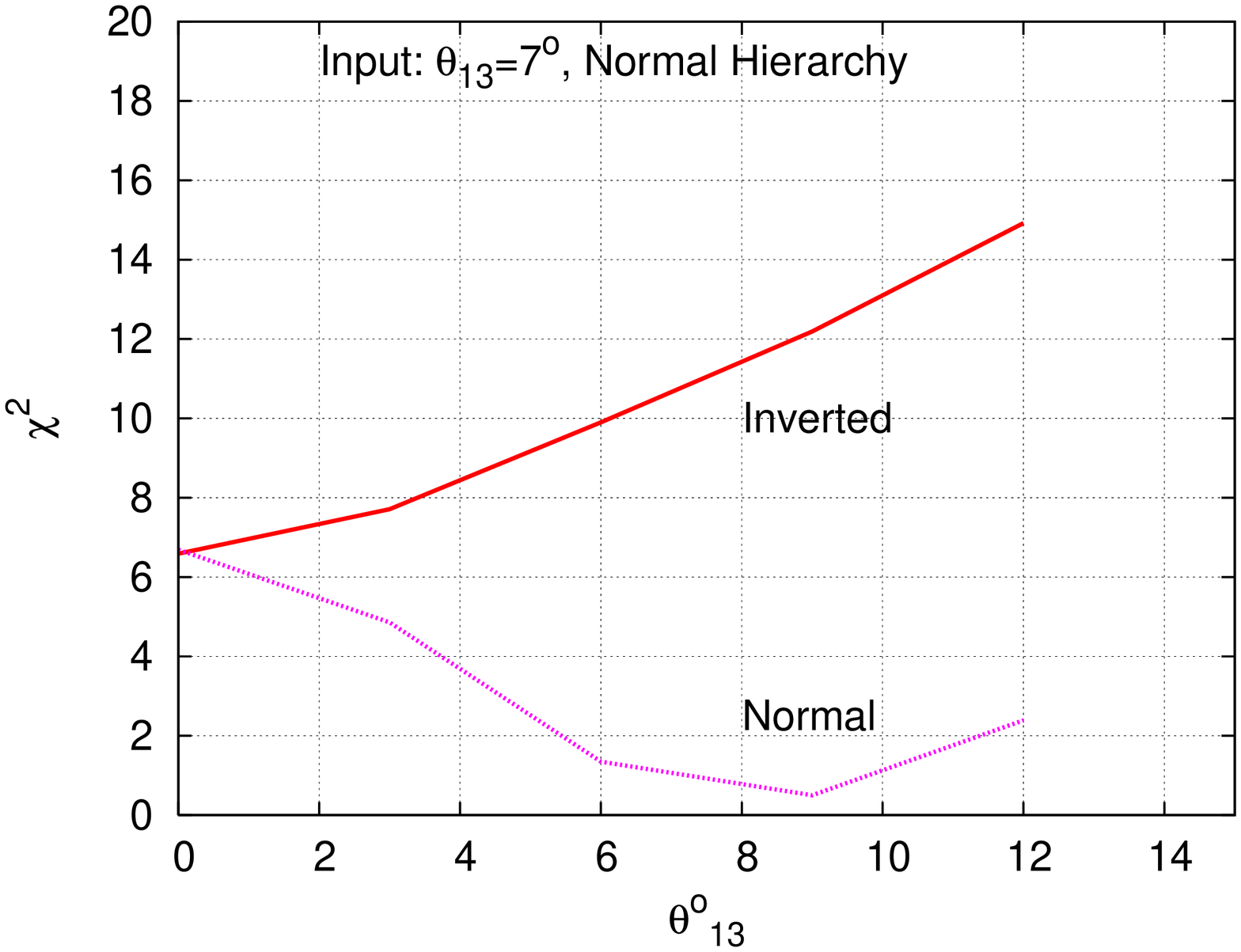}
\includegraphics[width=5.0cm,angle=270]{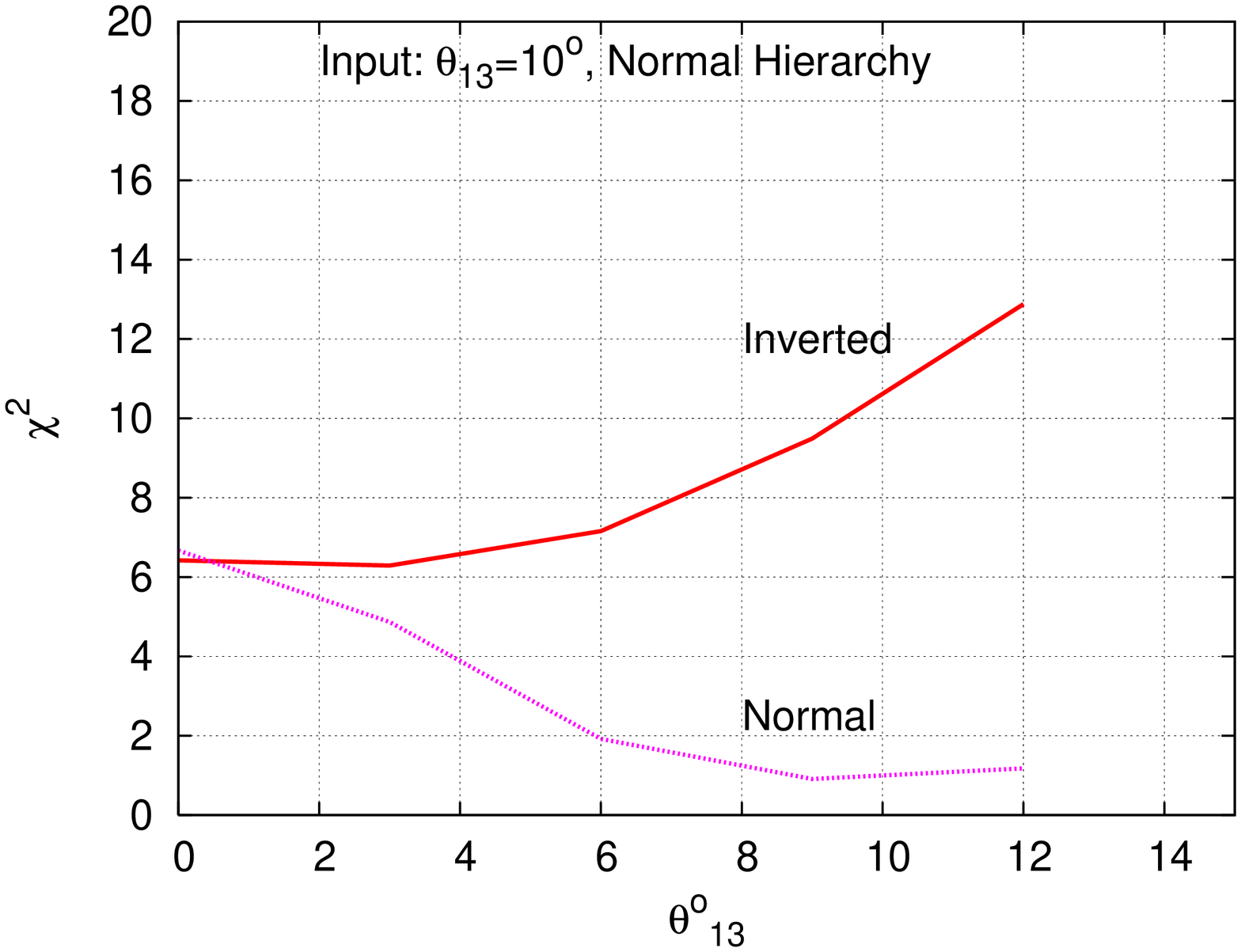}
\caption{\sf The same plots of fig. \ref{f:marginmt13}, but with
input of NH and $\theta_{13}$=$7^\circ$ (left) and
$10^\circ$ (right).}
\label{f:marginpt13}
\end{figure*}
\begin{figure*}[htb]
\includegraphics[width=5.0cm,angle=270]{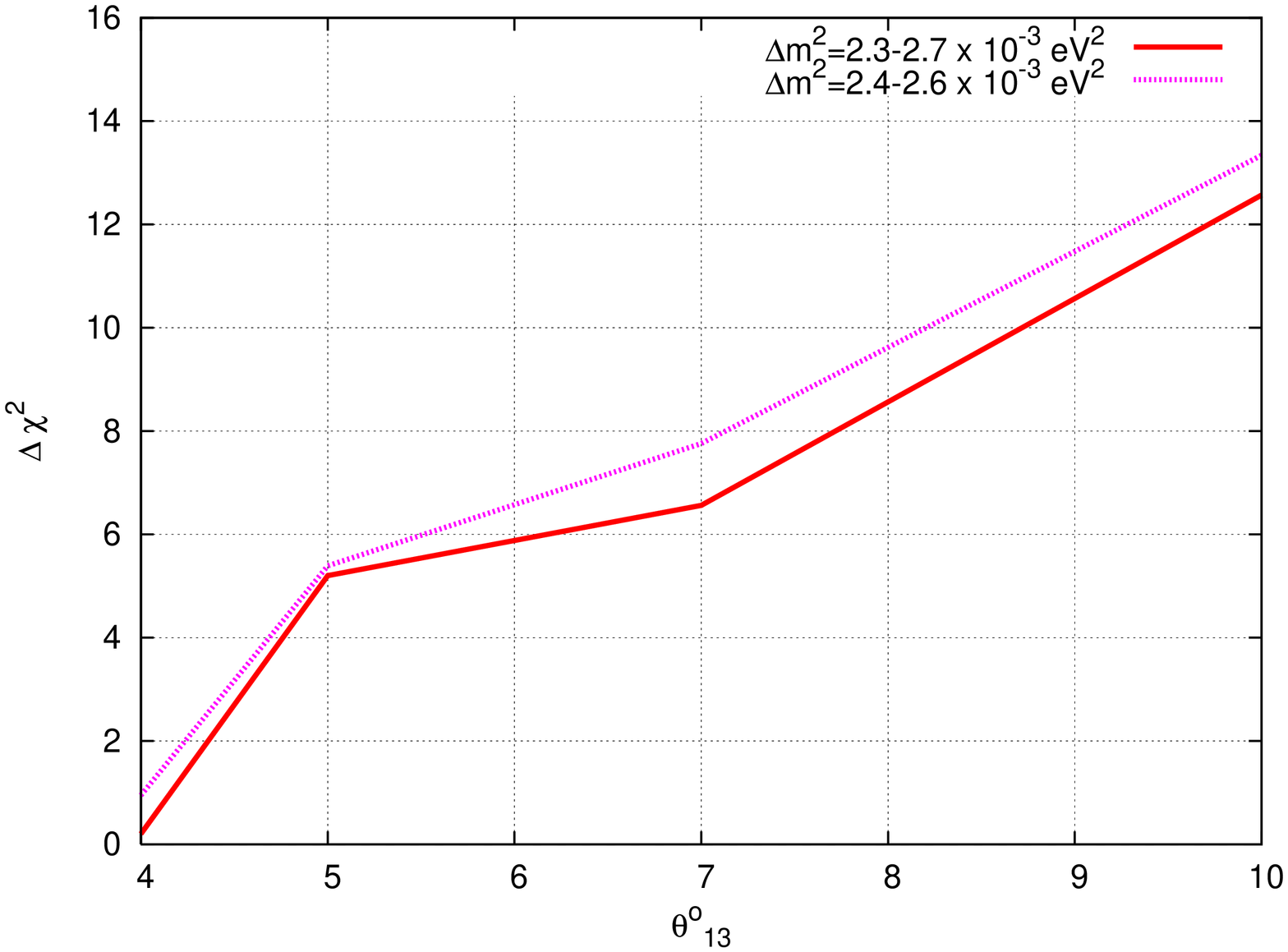}
\caption{\sf The variation of the discrimination sensitivity 
$\Delta \chi^2$
with
the input value of $\theta_{13} $ 
for two marginalization ranges of $|\Delta m_{32}^2|$.
The range of $\theta_{23}$ is  $39^\circ$ - $51^\circ$.
The type of input hierarchy is inverted. }
\label{f:margin}
\end{figure*}

\section{Discussion}\label{s:discussion}

Here  we will discuss some  points for the  estimation
of detector efficiency and 
the errors that may occur
in event reconstruction of the experimental data.

\begin{enumerate}
\item One can estimate the  efficiency of the  ICAL detector
in the following procedure.
The fiducial volume of the detector can be estimated so that 
an event with vertex inside this volume can be reconstructed.
To reconstruct an event one may need the minimum number of hits
$\gapp 9$. Therefore, one can  estimate the efficiency just neglecting 
10 layers from top and bottom and 50 cms from lateral sides.
For this into consideration, the efficiency may be roughly  
$$\approx \frac{120~{\rm (layers)}\times 4700~{\rm cm}\times 1500~{\rm cm}}
{140~{\rm (layers)}\times 4800~ {\rm cm}\times 1600~{\rm cm}}
\simeq 79 \%.$$  
\item 
For the maximum difference between NH and IH,  
we find the lowest value 
of $E_\mu \approx 1$ GeV at $\cosz_\mu=-1$   and the highest value of 
$E_\mu \approx 15$ GeV 
at  $\cosz_\mu\approx -0.3$. For these cases,  we will get number of hits
$\gapp 12$, and there will be  a very high efficiency of  filtration and reconstruction
of 
these events in actual experiment. 

\item 
For the muons below the energy of 15 GeV,
the charge identification
capability of ICAL@INO
is $\gapp 95\%$ with a magnetic field of 1 Tesla\cite{ino}.
Here, we consider only the muons
whose reconstruction  efficiency
is very high and the  
energy and angle resolutions
are within 3-6\% for the considered ranges.
Moreover, the wrong charge identification possibility is very 
negligible \cite{ino}. 
{\sf So one may easily expect that the result estimated in this paper will 
not change  appreciably after GEANT\cite{geant} simulation.}

\end{enumerate}

\section{Conclusion} 
{We have studied the neutrino mass hierarchy at the magnetized ICAL detector
 at INO 
with atmospheric neutrino events generated by the  Monte Carlo event generator
Nuance(version-3). We have done the analysis in muon energy and zenith angle bins.
We have adopted a numerical technique for the migration of number of neutrino
events from neutrino energy and zenith angle bins to muon energy and zenith 
angle bins to find the ``theoretical data" in chi-square analysis. The
so called ``experimental data" are obtained in muon energy and zenith angle
bins  from
the Nuance simulation.  The resonance ranges in terms of muon energy and zenith
angle are found for different values of $\theta_{13}$. Then we find  
suitable variables 
and make a marginalized chi-square analysis using the pull method
 for both true and false hierarchy.  
The choice of variables and the pull method  minimize the possible systematic 
uncertainties in the analysis.
Finally we find the sensitivity
as the difference $\Delta \chi^2 = \chi^2 {(\rm false)} - \chi^2 {(\rm true)}$
as a function of $\theta_{13}$ as shown in fig. \ref{f:margin}.
    }
It is found that the neutrino mass hierarchy with atmospheric neutrinos can be 
probed up to a low value of $\theta_{13}\approx 5^\circ$ at ICAL@INO
by judicious selection of events and observables. Here we have also shown 
that $>$ 95\% 
CL can
be achieved for discrimination of hierarchy at $\theta_{13}\gapp 5^\circ$
in a 1 Mton-year  exposure of ICAL detector.


\end{document}